\PassOptionsToPackage{unicode}{hyperref}
\PassOptionsToPackage{hyphens}{url}
\documentclass[
]{article}
\usepackage{amsmath,amssymb}
\usepackage{iftex}
\ifPDFTeX
  \usepackage[T1]{fontenc}
  \usepackage[utf8]{inputenc}
  \usepackage{textcomp} 
\else 
  \usepackage{unicode-math} 
  \defaultfontfeatures{Scale=MatchLowercase}
  \defaultfontfeatures[\rmfamily]{Ligatures=TeX,Scale=1}
\fi
\usepackage{lmodern}
\ifPDFTeX\else
\fi
\IfFileExists{upquote.sty}{\usepackage{upquote}}{}
\IfFileExists{microtype.sty}{
  \usepackage[]{microtype}
  \UseMicrotypeSet[protrusion]{basicmath} 
}{}
\makeatletter
\@ifundefined{KOMAClassName}{
  \IfFileExists{parskip.sty}{%
    \usepackage{parskip}
  }{
    \setlength{\parindent}{0pt}
    \setlength{\parskip}{6pt plus 2pt minus 1pt}}
}{
  \KOMAoptions{parskip=half}}
\makeatother
\usepackage{xcolor}
\usepackage[left=4cm,right=4cm,top=2cm,bottom=2cm]{geometry}
\usepackage{graphicx}
\makeatletter
\def\maxwidth{\ifdim\Gin@nat@width>\linewidth\linewidth\else\Gin@nat@width\fi}
\def\maxheight{\ifdim\Gin@nat@height>\textheight\textheight\else\Gin@nat@height\fi}
\makeatother
\setkeys{Gin}{width=\maxwidth,height=\maxheight,keepaspectratio}
\makeatletter
\def\fps@figure{htbp}
\makeatother
\providecommand{\tightlist}{%
  \setlength{\itemsep}{0pt}\setlength{\parskip}{0pt}}
\newlength{\cslhangindent}
\newlength{\csllabelwidth}
\newlength{\cslentryspacingunit} 
\newenvironment{CSLReferences}[2] 
 {
  \setlength{\parindent}{0pt}
  \ifodd #1
  \let\oldpar\par
  \def\par{\hangindent=\cslhangindent\oldpar}
  \fi
  \setlength{\parskip}{#2\cslentryspacingunit}
 }%
 {}
\usepackage{calc}

\usepackage{float} \usepackage{booktabs}  \usepackage{setspace}
\usepackage{booktabs}
\usepackage{longtable}
\usepackage{array}
\usepackage{multirow}
\usepackage{wrapfig}
\usepackage{float}
\usepackage{colortbl}
\usepackage{pdflscape}
\usepackage{tabu}
\usepackage{threeparttable}
\usepackage{threeparttablex}
\usepackage[normalem]{ulem}
\usepackage{makecell}
\usepackage{xcolor}
\ifLuaTeX
  \usepackage{selnolig}  
\fi
\IfFileExists{bookmark.sty}{\usepackage{bookmark}}{\usepackage{hyperref}}
\IfFileExists{xurl.sty}{\usepackage{xurl}}{} 
\hypersetup{
  pdftitle={Performance, Knowledge Acquisition and Satisfaction in Self-selected Groups: Evidence from a Classroom Field Experiment},
  pdfauthor={Julius Düker; Alexander Rieber},
  pdfkeywords={field experiment, group formation, self-selected groups},
  hidelinks,
  pdfcreator={LaTeX via pandoc}}

\title{Performance, Knowledge Acquisition and Satisfaction in
Self-selected Groups: Evidence from a Classroom Field Experiment}
\author{Julius Düker\footnote{Ulm University,
  \href{mailto:julius.dueker@uni-ulm.de}{\nolinkurl{julius.dueker@uni-ulm.de}}} \and Alexander
Rieber\footnote{Ulm University,
  \href{mailto:alexander.rieber@uni-ulm.de}{\nolinkurl{alexander.rieber@uni-ulm.de}}}}
\date{}

\begin{document}
\maketitle
\begin{abstract}
We investigate how to efficiently set up work groups to boost group
productivity, individual satisfaction, and learning. Therefore, we
conduct a natural field experiment in a compulsory undergraduate course
and study differences between self-selected and randomly assigned
groups. We find that self-selected groups perform significantly worse on
group assignments. Yet, students in self-selected groups learn more and
are more satisfied than those in randomly assigned groups. The effect of
allowing students to pick group members dominates the effect of
different group compositions in self-selected groups: When controlling
for the skill, gender, and home region composition of groups, the
differences between self-selected and randomly formed groups persist
almost unaltered. The distribution of GitHub commits per group reveals
that the better average performance of randomly assigned groups is
mainly driven by highly skilled individuals distributed over more groups
due to the assignment mechanism. Moreover, these highly skilled
individuals contribute more to the group in randomly formed groups. We
argue that this mechanism explains why self-selected groups perform
worse on the projects but acquire more knowledge than randomly formed
groups. These findings are relevant for setting up workgroups in
academic, business, and governmental organizations when tasks are not
constrained to the skill set of specific individuals.
\end{abstract}

\section{Introduction}

Many economic and social activities require teamwork. How well a team
performs depends not only on the individual characteristics and skills
of its members but also on how they interact and cooperate with each
other.\footnote{See, for instance, Dahl, Kotsadam, and Rooth (2021),
  Weidmann and Deming (2021), or Ai et al. (2023)} One factor that may
influence team cooperation is how the team is formed. If team members
can choose their partners, they may behave more altruistically or
reciprocally than if they are randomly assigned to a team.\footnote{See,
  e.g., Coricelli, Fehr, and Fellner (2004)} We find in a classroom
field experiment, that self-selected groups perform significantly worse
on group assignments. Still, students in these groups learn more and are
more satisfied than those in randomly assigned groups.

This paper examines how the process of group formation affects the
outcomes of teams and individuals in a cognitively demanding task. We
conduct a classroom field experiment in a data analysis course involving
group work and individual assessment. We assign students to either
self-select their teammates or be randomly matched with other students
to form groups of three. We follow the same students for two consecutive
semesters, switching the group formation method between semesters. We
complement our experimental data with administrative data on student
characteristics, which allows us to isolate the effects of group
composition and group formation on group performance, individual
learning, and satisfaction.

We measure group performance with the grades from the three data science
projects each group completes during a semester. These projects account
for 70\% of the final grade and require groups to analyze real economic
or corporate data, visualize data, and interpret results. We obtain our
measure of individual learning from the final exam at the end of the
semester, which accounts for 30\% of the final grade. We compute
satisfaction levels from surveys we require students to complete after
each project. Further, we include administrative data on students' high
school GPA as a proxy for individual ability and the place where a
student went to high school as a proxy for their geographic origin.
Additionally, we track individual contributions to the group projects
using time-stamped commits in (private) GitHub repositories.

We find that self-selected groups perform significantly worse on group
projects than randomly formed groups, but students in self-selected
groups learn more and are more satisfied than those in randomly formed
groups. We compare the effects of group composition and group formation
on group performance, individual learning, and satisfaction. Group
formation has a stronger impact than group composition on all three
outcome variables. Allowing students to choose their group members
increases their learning and satisfaction but reduces their performance,
holding group composition constant. However, self-selected groups also
have a different composition than randomly formed groups, which tends to
lower their performance, learning, and satisfaction. Self-selected
groups are more homogeneous in terms of GPA, gender, and geographic
origin. These findings are consistent with the literature on homophily
in group formation, e.g., McPherson, Smith-Lovin, and Cook (2001),
Carrell, Sacerdote, and West (2013), or Charroin, Fortin, and Villeval
(2022).

Combining both effects, we find that self-selected groups perform about
5.1 percentage points worse on group projects than randomly formed
groups. Moreover, the performance of self-selected groups is more
dispersed, reflecting the distribution of high school GPAs among groups.
However, individuals in self-selected groups learn more, scoring 3.3
percentage points higher on the individual exam. They also perceive
their group as more effective and report higher overall satisfaction
with their group, by 12.4 percentage points, than those in randomly
formed groups. To understand why self-selected groups perform worse on
group projects but better on individual exams, we examine the
distribution of work and skills within groups using GitHub data.

Using GitHub allows us to analyze the mechanisms behind these average
effects based on individual contributions to each project.\footnote{Isomöttönen
  and Cochez (2014) and Haaranen and Lehtinen (2015) present case
  studies on how to use GitHub and GitLab in classroom settings and
  examine whether a project-based course such as ours can teach students
  to use the platform correctly. See also Feliciano, Storey, and
  Zagalsky (2016) and Lu et al. (2017) for a review on how to use GitHub
  in the classroom.} Our analysis reveals that high-skilled students
(high GPA) contribute most of the code and text in randomly formed
groups. Self-selected groups, in contrast, distribute the workload more
evenly across skill levels.\footnote{Note, self-selected and randomly
  formed groups do not differ significantly in how equally they
  distribute work.} High-skilled and low-skilled students tend to
cluster in self-selected groups, while they are mixed in randomly formed
groups. As a result, randomly formed groups perform better in group
projects because high-skilled students do more work. However, this work
distribution also encourages low-skilled students to free-ride in
randomly formed groups, which impedes their individual learning.
Furthermore, the higher workload of high-contributing students in
randomly formed groups leads to lower satisfaction rates among these
students, which we cannot observe for self-selected groups.\footnote{Knez
  and Simester (2001), Bandiera, Barankay, and Rasul (2013) and De
  Paola, Gioia, and Scoppa (2019) argue that workers internalize the
  effects of their effort on coworkers when they feel socially connected
  to coworkers, which could result in less free riding and better
  performance.}

This paper adds to the literature on group formation and performance by
conducting a field experiment in a data analysis course that involved
high-stakes and cognitively challenging tasks. The most related studies
to ours are from Fischer, Rilke, and Yurtoglu (2023), Fenoll and
Zaccagni (2022), and Kiessling, Radbruch, and Schaube (2022), who also
conduct classroom field experiments to compare the performance of
self-selected and randomly formed groups.\footnote{See also Boss et al.
  (2021) and Chen and Gong (2018) for more classroom field experiments
  with self-selected groups} Fischer, Rilke, and Yurtoglu (2023) examine
how the group formation process affects the skill composition and
performance of groups. They find that self-selected groups are more
assortatively matched and perform similarly or worse than randomly
assigned groups in different tasks. Their task was low-stakes (15\% of
student grade) for a team of two, while ours was high-stakes (70\% of
final grade) for a group of three. While Fischer, Rilke, and Yurtoglu
(2023) find evidence that self-selected groups perform worse in a
classroom setting, Fenoll and Zaccagni (2022) and Kiessling, Radbruch,
and Schaube (2022) find evidence that the opposite is the
case.\footnote{Also Chen and Gong (2018) and Boss et al. (2021) find
  evidence that self-selected groups outperform randomly formed groups,
  but do not control for skill composition.} In Fenoll and Zaccagni
(2022), high school students are randomly or self-selected into groups
of six students during a mathematics summer camp. They find that
self-selected groups perform significantly better than randomly formed
groups in a competition. Kiessling, Radbruch, and Schaube (2022) show
that students perform significantly better in a running task when they
can self-select their running partner. They show that the group
formation process itself affects running performance, i.e., individuals
increase effort if they can choose their teammates. Kiessling, Radbruch,
and Schaube (2022) provide evidence for a positive formation effect in
self-selected groups and evidence that working with a friend or a
similar performer improves productivity.

We extend these studies in four ways: First, we examine a collaborative,
high-stakes task for a group of three students. This allows us to study
group dynamics that may differ from those in smaller groups. Second, we
measure not only group performance but also individual learning and
satisfaction, which we consider essential for a successful and lasting
collaboration in a group. Third, we combine our experimental data with
administrative data on student characteristics, which enables us to
separate the effects of group composition and group formation on the
outcomes. Fourth, we analyze individual contributions to the group,
which helps us to understand how groups allocate work among their
members.

The rest of the paper is organized as follows: In section two, we
describe our experimental setup, develop our hypothesis in section
three, and detail our empirical strategy in section four. In section
five, we describe our data and provide some descriptive statistics. In
section six, we present our main results on project performance,
individual knowledge gain, and satisfaction in self-selected and
randomly formed groups. In section seven, we explore the mechanisms
behind our results using our GitHub data on individual commits. In
section eight, we check the robustness of our main results. In the last
section, we conclude.

\section{Classroom Field Experiment}

\subsection{Background}

We conduct a field experiment in a classroom setting. The participants
are undergraduate students in management and economics at a German
university. They enroll in a compulsory data science course that covers
data analysis, reproducibility, R programming, statistical inference and
causality. The course lasts two semesters and students take it in their
second year. They get a separate grade for each semester and they need
to pass the first part to continue to the second part. The course
structure is as follows:

\vspace{2cm}

\begin{itemize}
\tightlist
\item
  \emph{Lectures:} In the first four weeks of each semester, the
  instructor introduces theoretical concepts on data science methods.
  The winter semester focuses on descriptive analysis and the summer
  semester on causal inference. The course materials are the same every
  year.\footnote{The (German) website of the course with all the
    material covered can be found here:
    \url{https://projektkurs-data-science-ulm2021.netlify.app/}}
\item
  \emph{Projects:} In the next 11 weeks of each semester, the students
  work in groups of three on three projects.\footnote{Besides the group
    projects, each student has to complete and pass six individual
    assignments in each semester: three interactive problem sets, an
    online test exam, and two review reports on other groups' projects.
    They need to score at least 80\% on each problem set and 30\% on the
    test exam to join a group. They need to write review reports to take
    the final exam. The students rank the review reports they receive
    from other groups and the instructors evaluate their quality
    (clearly structured, at least one point suggested for improvement,
    constructive). During the experiment, two students failed the test
    exam and we excluded them from our analysis. All other students
    passed these requirements and were admitted to the final exam.}
\item
  \emph{Grades:} In each semester, the students can earn up to 100
  points for their final grade. They get 10 points from a test project,
  30 points from each of the two main projects, and 30 points from a
  final exam with 30 multiple-choice questions. The test projects are
  identical for both cohorts and most of their content is covered in
  class. The main projects change every semester.
\end{itemize}

\subsection{Group Projects}

The group projects involve various types of economic analysis. The first
step is to acquire data from different sources, such as APIs, databases,
or web scraping. The following steps are to wrangle, describe, and
visualize the data. The projects in the second semester include
regression analysis and causal inference. Along with each project, the
groups must submit a short screencast presenting their main results,
accounting for 30\% of the project points. The projects are managed and
submitted through Github, which allows us to track the frequency,
content, and timing of each student's contribution.\footnote{A potential
  challenge for this course is free-riding in groups. Students may need
  help dealing with non-cooperative group members. We present three
  escalation levels for resolving group conflicts at the beginning of
  each semester. The first level involves a meeting with the auxiliary
  lecturer, where group members voice their concerns, receive advice,
  and make verbal commitments. The second level involves a meeting with
  both instructors, where we assign specific tasks to the uncooperative
  group member and monitor their progress. The third level involves
  splitting the group and requiring individual work for the remaining
  projects. We only reveal this option in the final meeting, not
  earlier. In our two-year experiment, only one group reached the third
  level, two groups reached the second level, and three groups reached
  the first level.} After the deadline, we download the projects,
pseudonymize them, and randomly assign them to one of the instructors
for grading.\footnote{Note that pseudonyms vary for each group
  throughout a semester.} The instructors use a detailed rubric for each
question to ensure objectivity and consistency. We inform the groups
about their project points in the last week of the semester and at least
five days before the final exam.

We also ask student assistants to grade the final submissions of both
projects each semester on a 7-point Likert scale and provide feedback on
improving the projects. We give the student assistants instructions on
how to rate the projects and blind them to the group names and their
tutorial assignments.\footnote{Appendix A presents the instructions to
  student assistants before rating the projects. All student assistants
  have completed both parts of the course in previous years with a top
  quartile grade.} The groups receive feedback on improving their
projects from the student assistants a week after submitting their
projects.

\subsection{Intervention}

We want to compare the outcomes of self-selected and randomly formed
groups in a compulsory data science course. We conduct a natural field
experiment with two cohorts of students who take the course in two
consecutive years, cohort 2020 and cohort 2021. At the start of the
course, we ask students to consent to share their data on various
measures, such as GitHub commits, grades, high school grade point
average (GPA), quiz and online lab answers, project descriptions, and
feedback reports. We do not tell students about the experiment, but only
that we use their data to improve the course. The course grade mainly
depends on group work, which differs from previous studies using
low-stakes tasks.\footnote{E.g., Kiessling, Radbruch, and Schaube
  (2022), Fenoll and Zaccagni (2022), Fischer, Rilke, and Yurtoglu
  (2023), and Boss et al. (2021) use tasks accounting up to 15\% of the
  final grade.} Wise and DeMars (2005) and Ofek-Shanny (2020) show that
students provide significantly less effort with low stakes than high
stakes tasks.

Figure 1 shows the timeline and assignments for each cohort and each
course part. We explain them below.

\begin{figure}[H]

{\centering \includegraphics[width=1\linewidth]{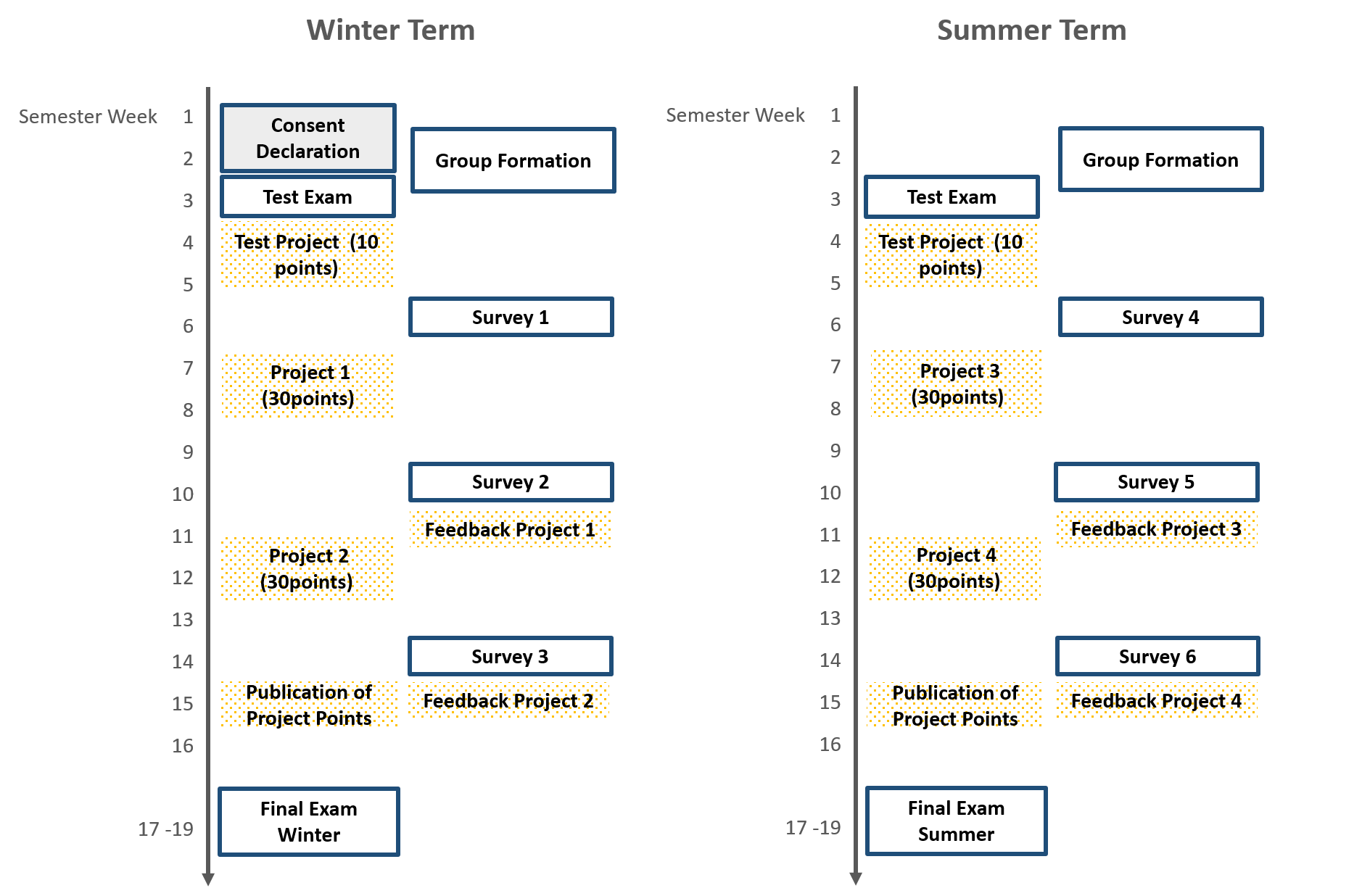} 

}

\caption{Timeline}\label{fig:unnamed-chunk-4}
\end{figure}

We compare two ways of forming groups of three students for the group
projects: self-select and random. In the self-select specification,
students have three weeks to create their groups using an online
learning platform. On the platform, they can observe the members of the
already formed groups at any time. We randomly assign the remaining
students to groups at the end of the third week. In the random
specification, we ask students if they want to participate in the group
projects and then randomly assign them to groups. We inform students
about their group members in week three.\footnote{When the number of
  participants does not allow for groups of three only, we form one,
  respectively two groups of two. We exclude these groups of two
  students from our analysis. Moreover, we exclude groups from the
  self-selection treatment that were formed by us, because their members
  failed to find a group.}

We survey students after each project via GitHub on their satisfaction
and perception of group collaboration.\footnote{See Appendix B for
  survey questions.} We assure students that their answers are
confidential and have no consequences on the course. We also include two
questions from the university course evaluation form to check the
validity and seriousness of our survey.\footnote{The answers are
  consistent, e.g., for the question ``In this course, I learn things
  that fill me with enthusiasm,'' we receive an average of 3.71 in our
  survey and an average of 3.79 in the university evaluation.}

We use a within-subjects design where all cohort students have the same
treatment during one semester. Table 1 shows how we distribute
treatments across cohorts and semesters. Students in cohort 2020
self-select groups in the winter term and are randomly assigned in the
summer term. Students in cohort 2021 are randomly assigned in the winter
term and self-select groups in the summer term. We registered a
pre-analysis plan in the American Economic Association RCT registry and
obtained IRB approval before the intervention.\footnote{See
  \url{https://doi.org/10.1257/rct.6726-1.0}}

\renewcommand{\arraystretch}{2}
\begin{table}

\caption{\label{tab:Number of students in each semester}Participants in Each Semester}
\centering
\fontsize{8}{10}\selectfont
\begin{tabu} to \linewidth {>{\raggedleft}X>{\raggedright}X>{\raggedright}X>{\raggedleft}X>{\raggedleft}X>{\raggedleft}X>{\raggedleft}X}
\toprule
Cohort & Term & Group Formation & Participants & In Group of Three & Declared Consent & Comply with Treatment\\
\midrule
2020 & winter & Self-Selected & 78 & 78 & 72 & 69\\
2020 & summer & Random & 63 & 63 & 57 & 57\\
2021 & winter & Random & 77 & 69 & 67 & 67\\
2021 & summer & Self-Selected & 73 & 66 & 57 & 52\\
\bottomrule
\end{tabu}
\end{table}

The Corona pandemic forced the university to switch to online teaching
from April 2020 to August 2021, covering the period of the field
experiment for cohort 2020. The pandemic affected the field experiment
in two ways: First, we delivered lectures and tutorials online for both
cohorts, using video tutorials and live streams. Second, students had
less in-person contact with each other, which could influence their
group formation. We account for this by implementing the self-select
treatments when students still knew each other from previous semesters
or had the opportunity to meet in person the semester before.
Additionally, we ask students in the surveys about their previous
relationships with their team members to analyze whether the pandemic
affected the formation of self-selected groups.

\section{Hypotheses Development}

Several factors play a role in determining individual learning and
satisfaction within a group, as well as group performance. These factors
can lead to significant performance differences between self-selected
groups and those formed randomly. In this paper, we aim to untangle the
effects that arise from the group formation process itself (referred to
as the ``formation effect'') from those that occur due to differences in
the group's composition (referred to as the ``composition effect''). We
pre-registered all our hypothesis in a pre-analyis plan before the start
of the experiment in November 2020.\footnote{See AEA RCT Registry:
  \url{https://doi.org/10.1257/rct.6726-1.0}. In the pre-analysis plan,
  we use the terms endogenous and exogenous instead of self-selected and
  random. From now on, we refer to endogenously formed groups as
  self-selected groups and exogenously formed groups as randomly formed
  groups to stress the group formation process. Compared to the
  pre-analysis plan we switch hypotheses 3 and 4.}

We conclude from the literature that group formation affects performance
positively in self-selected groups. Previous studies have shown that
self-selected groups exert more effort than randomly formed groups,
which could improve their performance (Kiessling, Radbruch, and Schaube
(2022), Coricelli, Fehr, and Fellner (2004)).

But, following the literature, group composition affects performance in
more nuanced ways. First, members in self-selected groups are more
similar to each other than in random groups in many dimensions (e.g.,
McPherson, Smith-Lovin, and Cook (2001), Charroin, Fortin, and Villeval
(2022), Fischer, Rilke, and Yurtoglu (2023)). This similarity can
increase group identity and coordination, as well as the internalization
of effort externalities (Bandiera, Barankay, and Rasul (2005), De Paola,
Gioia, and Scoppa (2019), Ai et al. (2023)). Second, self-selected
groups are more homogeneous regarding skill as individuals choose
teammates with similar backgrounds and preferences (e.g., Charroin,
Fortin, and Villeval (2022) and Ruef, Aldrich, and Carter (2003)).
Fischer, Rilke, and Yurtoglu (2023) also finds such a selection of group
members with comparable skill sets in an undergraduate classroom
setting. However, this homogeneity can reduce skill spillovers within a
group and, therefore, lower performance (e.g., Page (2007), Mas and
Moretti (2009), or Hamilton, Nickerson, and Owan (2003)). The type of
task also matters for group performance, especially for more vs.~less
cooperative tasks (e.g., Fischer, Rilke, and Yurtoglu (2023)). Following
Fischer, Rilke, and Yurtoglu (2023), we assume skill spillovers outweigh
effort externality internalization and higher group identity in
homogeneous groups. Therefore, we expect that group composition affects
performance negatively in self-selected groups.

We do not know which effect is stronger: group formation or group
composition. Group formation implies that self-selected groups perform
better, while group composition implies that randomly formed groups
perform better. Therefore, our first hypothesis is undirected:

\emph{H1: On average self-selected and randomly formed groups do not
differ in terms of project points.}

We also want to measure the formation and composition effects
separately. We use the following hypothesis to test if there is a
formation effect in favor of self-selected groups\footnote{We also
  pre-registered sub-hypotheses for each main hypothesis where we wanted
  to analyze the impact of heterogeneity within groups. But because of
  insufficient statistical power we can not test hypotheses H1b, H2b,
  H3b, and H4b from the pre-analysis plan and thus do not include them
  in this section.}:

\emph{H1a: Groups with similar skill composition perform better under
self-selection treatment compared to random formation treatment}

Group composition also affects individual learning in a group. We
conjecture that skill spillovers are the primary driver of group
learning (See also Page (2007), Mas and Moretti (2009), or Hamilton,
Nickerson, and Owan (2003)). The more skill-diverse a group is, the more
skill spillovers occur between group members and the more they learn.
This might be offset by other dimensions of similarity (e.g., age,
gender, or home region) in self-selected groups, but we find no evidence
for that in the literature. Therefore, we expect that group composition
affects knowledge acquisition negatively in self-selected groups.

The formation effect might positively affect learning in self-selected
groups because group members provide more effort. However, we expect
this effect to be smaller than the composition effect from skill
spillovers. We, therefore, hypothesize that the overall effect of
self-selected group formation on knowledge acquisition is negative and
state the following hypothesis:

\emph{H2: On average the knowledge gain throughout a semester is larger
for randomly formed groups.}

We use the following hypothesis to test the formation effect on
knowledge acquisition:

\emph{H2a: Members of groups with similar skill composition learn more
under self-selection treatment compared to random formation treatment}

Moreover, we want to analyze individuals' satisfaction with the team. We
think high individual satisfaction with the team is essential for
long-lasting, efficient collaboration, as an individual's low
satisfaction can disperse and harm collaboration. Individual
satisfaction may be influenced by group formation and group composition.
We argue that individuals value being able to form groups themselves and
are more satisfied due to the formation effect in the self-selection
treatment.

We typically enjoy spending time with someone we can relate to more than
with strangers. Moreover, individuals prefer working with others of
their own kind, and we expect higher group identity and satisfaction in
homogenous groups than in heterogeneous groups. Since individuals tend
to form groups with others with similar skills and preferences
(McPherson, Smith-Lovin, and Cook (2001), Carrell, Sacerdote, and West
(2013), and Charroin, Fortin, and Villeval (2022)), we expect the
composition effect of self-selection on satisfaction, as well as the
overall effect on satisfaction to be positive. Thus, we state hypotheses
H3 and 3a as follows:

\emph{H3: Members of self-selected groups are more satisfied with the
group composition than those from randomly formed groups.}

\emph{H3a: Members of groups with similar skill composition are more
satisfied under self-selection treatment compared to random formation
treatment}

We argue above that self-selected groups are more homogenous,
internalize the externalities of their own effort provision, and
identify themselves stronger with the group. For this reason, group
members might contribute more to the group, particularly those who would
otherwise contribute little. Low contributors might increase their
effort provision to a larger extent, because it is easier for them
coming from a low effort level. Therefore, we assume that self-selected
groups distribute work more equally due to the composition effect.

Similar to the hypotheses above, we expect the formation effect to be in
favor of self-selected groups. Individuals provide more effort when they
can choose who to work with. This might induce a more equal distribution
of work, as it is easier for low contributors to increase their effort
provision. Hence, when taking composition and formation effect together,
we also expect that self-selected groups distribute work more equally.
This is in line with (Hamilton, Nickerson, and Owan (2003)). Thus, we
state the following two hypothesis:

\emph{H4: Members of self-selected groups contribute more equally to the
project.}

\emph{H4a: Members of groups with similar skill composition contribute
more equally under self-selection treatment compared to random formation
treatment}

\section{Empirical Strategy}

We define the composition effect as the differences in group
performance, individual knowledge acquisition, and satisfaction between
self-selected and randomly formed groups, which we can explain with home
region, skill, or gender composition. In contrast, the formation effect
describes the differences between self-selected and randomly formed
groups, which we cannot attribute to a distinct home region, skill, or
gender composition.\footnote{We cannot distinguish whether the formation
  effect arises from giving students the choice to select their group
  members or from working with someone they have a connection with. This
  is a difficult question to answer, since self-selected groups tend to
  consist of friends or acquaintances, while randomly formed groups do
  not. Even when random groups include friends, they are likely to be
  less close than self-selected groups. This is also true in real-world
  settings, where self-selection often implies some degree of social
  ties.}

We estimate the effect of the formation mechanism on group performance
using the following specification:

\begin{equation}
y_{jp} = \beta_1 S_{ct} + \beta_2 X_{j} + \gamma_{p} + \delta_{t} +  \epsilon_{jp}
\end{equation}

where \(y_{jp}\) is the performance of group \(j\) on project \(p\),
\(S_{ct}\) is an indicator variable for self-select group formation in
cohort \(c\) and term \(t\). \(X_{j}\) is a vector of group covariates,
including skill composition, geographic origin composition, and gender
of group members. \(\gamma_{p}\) represent project fixed effects,
\(\delta_{t}\) are term fixed effects, and \(\epsilon_{jp}\) is the
error term. The coefficient \(\beta_1\) captures the causal effect of
self-select group formation on group performance. We cluster standard
errors at the group level to account for within-group correlation.

In our regression specification, the composition effect is the
difference in \(S_{ct}\) between the baseline model, not including any
control variables, to the full model with all control variables
\(X_{j}\).

We measure group performance with the student assistant grades from the
group projects. Student assistants are unaware of the experiment and
anonymously grade group projects on a 1 to 7 scale. We tell student
assistants that their grading has no impact on the final grade of the
groups they rate and remunerate them to evaluate the projects.
Additionally, for the student assistant grading, we pseudonymize group
names.\footnote{Student assistants grade projects and screencast
  separately. For our analysis, we calculate a weighted mean from these
  two grades. We use the same 70/30 weights we use for the lecturer
  grades.} We linearly transform these grades to a 0 to 100 scale for
better interpretability and comparability. Our main proxy for prior
skill is the high school GPA of a student. For our skill composition
covariates, we compute the difference between the group members with the
best and the worst high school GPAs of a group and take the high school
GPAs of the best and the second-best members of a group. To determine
the geographic origin of students, we use the county where they
graduated from high school. In our regression analysis, we include a
dummy variable to indicate whether at least two group members originate
from the same or neighboring counties. We ascertain students' genders
from their first names.

We estimate the effect of the formation mechanism on individual
knowledge acquisition and satisfaction using the following
specification:

\begin{equation}
a_{itc} = \beta_1 S_{ct} + \beta_2 X_{j} + \delta_{t} + \epsilon_{itc}
\end{equation}

where the dependent variable \(a_{itc}\) is knowledge acquisition of
individual \(i\) from cohort \(c\) in term \(t\). We measure knowledge
acquisition as the share of correct answers in the final multiple-choice
exam at the end of each semester, concentrated on the 15 questions about
the projects. We use the same specification to estimate the effect of
group formation on satisfaction. We obtain satisfaction levels from the
surveys after each project on a scale from 1 to 5. We average the
satisfaction measure regarding the two group projects of each semester
for each student and then transform them to a scale from 0 to 100.

We estimate the effect of the formation mechanism on the distribution of
individual contributions within a project using the following
specification:

\begin{equation}
g_{ip} = \beta_1 S_{ct} + \beta_2 X_{j} + \delta_{t} +  \epsilon_{ip}
\end{equation}

where \(g_{ip}\) is the individual contribution of student \(i\) on
project \(p\). To assess an individual's contribution to a project, we
first calculate the total number of words of code and text contributed
by that individual and the total number of words of code and text
contributed by the entire group to the project. The share of words an
individual contributes to the total amount of words contributed to a
project then gives us our measure for individual contribution. Since we
are interested in how groups distribute work, we calculate the standard
deviation of the share of words contributed for each group and project.
We then use this standard deviation as the dependent variable for the
regression above.

\section{Data and Descriptives}

Our data includes information about group performance, individual
contributions to projects, satisfaction, and knowledge acquisition over
time. We track individual contributions on GitHub through commits made
by each student. These commits allow us a) to compute how many words a
student committed to a project, b) if they committed code and text to
the project, and c) which questions they answered with that commit.
Further, we implemented a survey after each project, which allows us to
measure individual satisfaction and perceived contributions on a 5-point
likert-scale.\footnote{In Appendix B, we provide all questions we asked
  in this survey.} We can measure individual knowledge acquisition using
the exam results administered at the end of each semester.

All students from cohorts 2020 and 2021 who took part in the compulsory
data science course described in the ``Background'' section are eligible
to take part in our field experiment, given they participated in the
project phase. We exclude all student groups consisting of two or four
members from our analysis.\footnote{Some groups have two or four members
  if the number of students in a term is not divisible by three.} We
also exclude individuals who do not consent to have their data used.
Additionally, we exclude groups that were under the self-select
treatment formed by us randomly and consist of students who did not
manage to form a group themselves. Using these exclusion restrictions,
we arrive at a total of 138 students in our data. Most students received
both treatments: 121 participated in the self-select treatment and 124
in the random formation treatment. Table 1 depicts the number of
students participating in our field experiment each semester. Albeit we
implement a within-subjects design, there are differences in the number
of students in each term within a cohort. We observe attrition in our
sample when people leave the program after the first part of the course
or because students quit during a semester due to personal reasons. This
could also result in a higher or lower number of groups with two or four
members, which we exclude from our analysis. We show in Appendix C, with
a balance table for the subject-specific covariates, that there are no
significant pre-experimental differences between students in
self-selected and randomly-formed groups.

\begin{table}

\caption{\label{tab:GPADistribution}Summary Statistics of Main Variables}
\centering
\fontsize{8}{10}\selectfont
\begin{tabu} to \linewidth {>{\raggedright\arraybackslash}p{4cm}>{\raggedright}X>{\raggedleft}X>{\raggedleft}X>{\raggedright}X>{\raggedleft}X>{\raggedleft}X>{\raggedleft}X}
\toprule
Variable & Obs. Unit & Mean & N & SD & Min & Median & Max\\
\midrule
Gender: Female & Student & 0.35 & 138 &  & 0.00 & 0.00 & 1.0\\
High School GPA & Student & 2.33 & 138 & 0.63 & 1.00 & 2.30 & 3.6\\
Two group members neighboring county & Student Term & 0.51 & 245 & 0.5 & 0.00 & 1.00 & 1.0\\
Exam Points & Student Term & 19.79 & 241 & 4.45 & 5.00 & 20.00 & 29.0\\
Exam Points Related to Projects & Student Term & 8.84 & 241 & 2.43 & 2.00 & 9.00 & 14.0\\
\addlinespace
Satisfaction with Team & Student Term & 4.09 & 241 & 1.01 & 1.00 & 4.00 & 5.0\\
Perceived contribution & Student Project & 0.38 & 490 & 0.11 & 0.04 & 0.35 & 0.8\\
GitHub Committed Words & Student Project & 2253.36 & 490 & 1784.02 & 0.00 & 1902.00 & 8750.0\\
Student Assistant Project Rating & Group Project & 5.64 & 172 & 1.09 & 3.00 & 6.00 & 7.0\\
\bottomrule
\end{tabu}
\end{table}
\renewcommand{\arraystretch}{1}

Table 2 presents descriptive statistics on the main dependent and
independent variables. The student assistants rated the projects on a
7-point Likert scale, averaging 5.64 points (SD=1.09). The students also
took a final exam worth 30 points and scored an average of 19.79 points
(SD=4.45), with a wide dispersion. In the exam questions related to the
projects, students scored an average of 8.84 points (SD=2.43). The
students reported their satisfaction with their teams on a 5-point
Likert scale, averaging 4.10 (SD=1.01). They also reported their
(perceived) contributions to the project, with an average of 0.38
(SD=.11), indicating a relatively equal work distribution among team
members. We measured the actual contributions by the number of words
each student committed to the project on GitHub, with an average of 2362
words (SD=1874), but with considerable variation across team members. We
also collected administrative data on the high school grade point
average (GPA) for all 138 participating students. The German GPA ranges
from 4.0 (the worst) to 1.0 (the best) and is the main criterion for
university admission in Germany and a good proxy for the general skill
level of students (e.g., Fischer and Kampkötter (2017)). The average GPA
in our sample is 2.33 (SD=.63). Finally, table 2 shows that 56\%
(SD=0.50) of the group members came from neighboring counties.

\section{Results}

In this section, we will shed light on how the group-forming mechanism
affects group composition, individual and group performance,
satisfaction, and knowledge acquisition. We aim to rule out the effect
of group composition on group performance, satisfaction, and knowledge
acquisition.

\subsection{Group Composition}

To be able to rule out the effect of group composition on individual and
group outcomes, we first need to examine whether and to which extent
self-selected and randomly formed groups differ in group composition.

\begin{figure}
\centering
\includegraphics{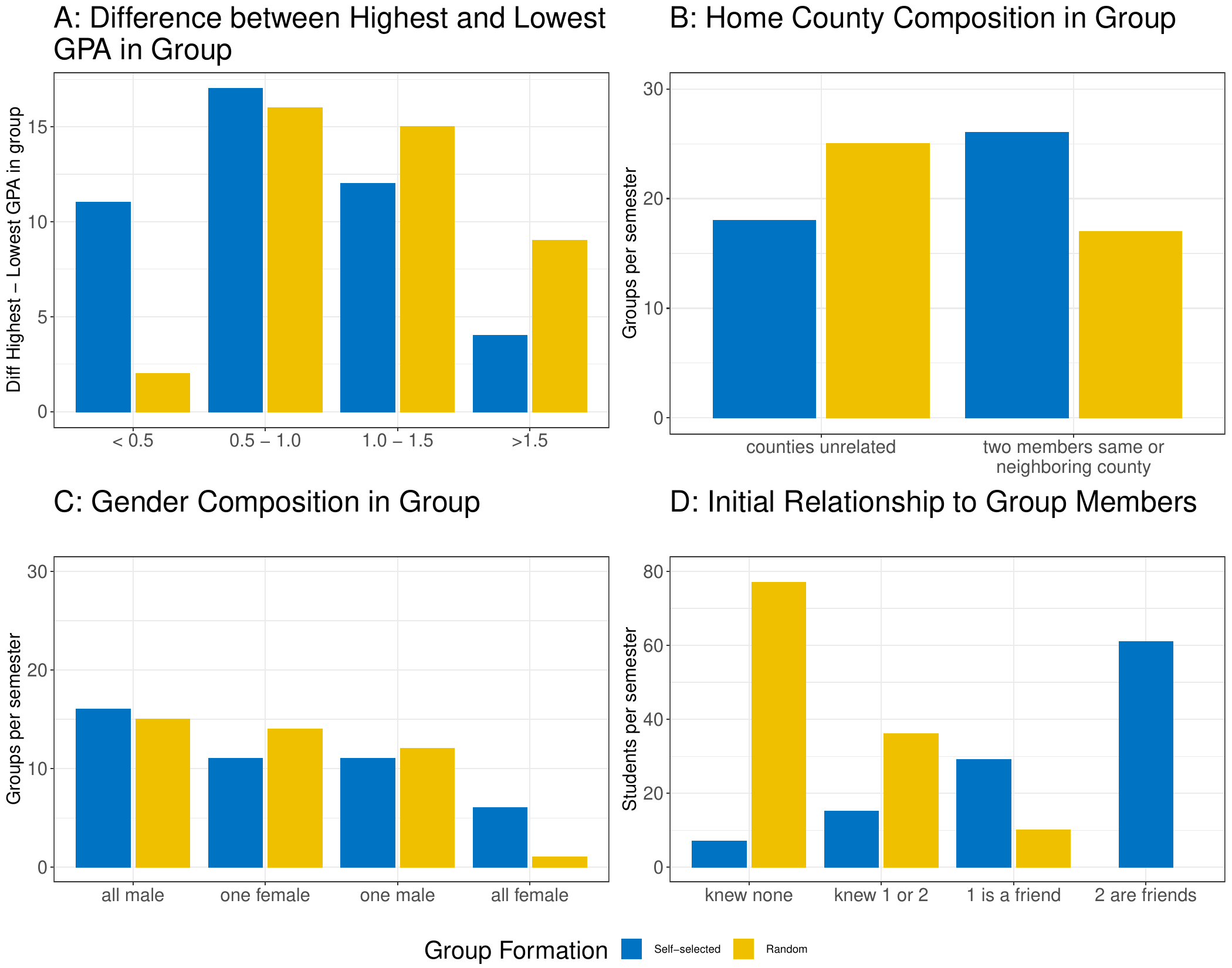}
\caption{Group Composition}
\end{figure}

\vspace{-1cm}

\begin{singlespacing}
Notes: We calculate the difference in high school GPA in a group between the members with the best and the worst high school GPA. Home counties correspond to the county in which a student graduated from high school. We gather information on initial relationships with the first survey. All panels but panel C show group numbers; in panel C, each observation is an individual.

\end{singlespacing}

Figure 2 shows the differences in four dimensions: skill, friendship,
home county, and gender. We use high school GPA as a proxy for skill and
the county where the student attended high school as a proxy for the
home county. In the first survey, we asked the students about their
friendship status at the beginning of the semester. We get the gender of
the students from their first names. In panel A of figure 2, each dot
represents a group in a semester, and the cross marks the mean of the
distribution. We also plot a boxplot and a kernel density distribution
in panel A. Figure 2 reveals that self-selected groups are more
homogenous in skill, more likely to come from the same or neighboring
counties, more likely to be of the same gender, and more likely to
include at least one friend than randomly formed groups. This is
consistent with the literature on homophily and the nature of random
sampling.

\subsection{Project Performance}

We are interested in how group performance is affected by self-selection
and by the characteristics of the group members. Figure 3 visualizes
differences in group performance between self-selected and randomly
formed groups.

\begin{figure}
\centering
\includegraphics{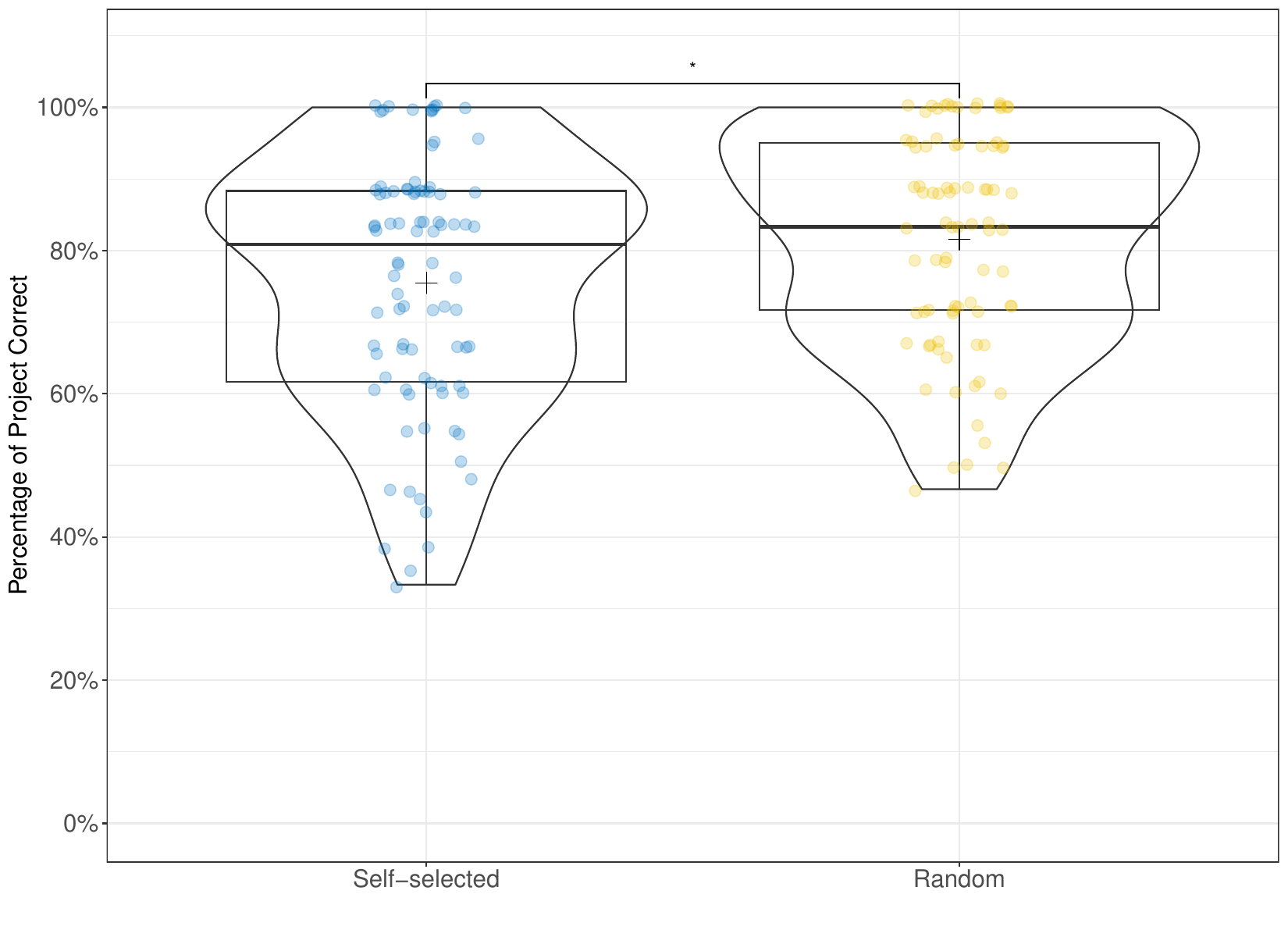}
\caption{Relationship between Group Performance and Treatment}
\end{figure}

\vspace{-1cm}

\begin{singlespacing}
Notes: Each dot represents a group and a project. We mark the mean with a cross. Further, we plot a kernel density and a boxplot. We measure project performance on a linear scale from 1 to 7, where 7 refers to the best performance. For our analysis, we transform this project performance measure to a 0 - 100 scale.* indicates that the p-value for the difference in means is below .1.

\end{singlespacing}

\begin{table}[H] \centering 
  \caption{Effect of group-forming mechanism on productivity} 
  \label{} 
\begin{tabular}{@{\extracolsep{-5pt}}lcccc} 
\\[-1.8ex]\hline 
\hline \\[-1.8ex] 
 & \multicolumn{4}{c}{\textit{Dependent variable:}} \\ 
\cline{2-5} 
\\[-1.8ex] & \multicolumn{4}{c}{Project Percentage Points} \\ 
\\[-1.8ex] & (1) & (2) & (3) & (4)\\ 
\hline \\[-1.8ex] 
 Self-selection & $-$5.134$^{**}$ & $-$4.384$^{*}$ & $-$4.226$^{*}$ & $-$6.970$^{**}$ \\ 
  & (2.418) & (2.297) & (2.410) & (2.925) \\ 
  & & & & \\ 
 Best Group GPA &  & $-$6.638 & $-$6.603 &  \\ 
  &  & (5.801) & (6.237) &  \\ 
  & & & & \\ 
 Second Best Group GPA &  & $-$3.287 & $-$2.957 &  \\ 
  &  & (4.150) & (4.232) &  \\ 
  & & & & \\ 
 Max GPA Difference &  & $-$0.556 & 0.326 &  \\ 
  &  & (3.792) & (4.102) &  \\ 
  & & & & \\ 
 All Members Female &  &  & $-$0.070 &  \\ 
  &  &  & (4.227) &  \\ 
  & & & & \\ 
 One Member Male &  &  & $-$0.293 &  \\ 
  &  &  & (3.143) &  \\ 
  & & & & \\ 
 All Members Male &  &  & $-$2.605 &  \\ 
  &  &  & (3.373) &  \\ 
  & & & & \\ 
 Two Members Same Region &  &  & 0.367 &  \\ 
  &  &  & (2.244) &  \\ 
  & & & & \\ 
 Group Mean Test Exam &  &  &  & 0.308$^{**}$ \\ 
  &  &  &  & (0.145) \\ 
  & & & & \\ 
 Self-selection:Summer Term &  &  &  & 3.285 \\ 
  &  &  &  & (4.878) \\ 
  & & & & \\ 
\hline \\[-1.8ex] 
Project FE & X & X & X &  \\ 
Term FE & X & X & X & X \\ 
Observations & 172 & 172 & 172 & 172 \\ 
R$^{2}$ & 0.075 & 0.161 & 0.168 & 0.079 \\ 
Adjusted R$^{2}$ & 0.052 & 0.125 & 0.110 & 0.057 \\ 
Residual Std. Error & 13.896 & 13.350 & 13.464 & 13.862 \\ 
\hline 
\hline \\[-1.8ex] 
\multicolumn{5}{l}{$^{*}$p$<$0.1; $^{**}$p$<$0.05; $^{***}$p$<$0.01} \\ 
\multicolumn{5}{l}{\parbox[t]{10cm}{Notes: Standard errors reported in parentheses are clustered at the group level. In the regressions, a group and a project define an observation. The dependent variable is the points awarded to the groups by the student assistants transformed to a 0 to 100 scale. Best Group GPA is a control variable for the best high school GPA in a group, and Second Best Group GPA for the second best high school GPA in a group. German high school GPAs range from 1 to 4, where 1 refers to the best grade and 4 to the worst. Max GPA Difference is a control variable for the difference in high school GPA between the member of the group with the best GPA and the one with the worst GPA. Two members same region is 1 if two members of a group graduated from high school in the same or neighboring counties, otherwise it is 0.}} \\ 
\end{tabular} 
\end{table}

Figure 3 suggests that randomly formed groups outperform self-selected
groups on the projects, with a smaller dispersion of project grades, and
contribute more to the projects. To receive an unbiased estimate of this
relationship and to be able to control for confounders, we run four OLS
regressions in table 3 on the effect of self-selected group formation on
group performance in the projects. The first regression in table 3
(column 1) gives us an overall effect: We compare the average
performance of self-selected and random groups without controlling for
group characteristics. To disentangle the formation and composition
effects, we add controls for the skill composition of a group in column
2. A higher project performance could result from knowledge spillovers
within the group, where the most skilled member instructs the other
group members on how to perform the task, or from one skilled group
member completing most of the project alone. Thus, in column 2, we
control for the skill level of the most skilled member of a group using
their high school GPA.\footnote{Note that German GPA is a linear scale,
  with 1.0 being the best grade and 4.0 the worst among students who
  receive a university entry certificate.} and for skill heterogeneity
within a group using the difference between the group's highest and
lowest high school GPA. Group identity may also affect collaboration
within a group. Ai et al. (2023) shows that individuals with a high
hometown similarity exert higher effort in their groups. Therefore, we
add in column 3 covariates on gender composition and the home region of
group members. Our pre-analysis plan pre-registered one regression to
study the formation and composition effect on group productivity. We
present the results in column 4. Unlike in columns 2 and 3, we proxy
individual skill levels with percentage points scored by a student in
the test exam.\footnote{We deviate in columns 2 and 3 from the
  pre-registered regressions by using high-school GPA minima and
  within-group differences for two reasons: First, when submitting the
  pre-analysis plan, it was not clear that we would obtain
  administrative data on high-school GPAs. Second, we were forced to let
  students do the test exam online during the Corona lockdown. In the
  online test exams, we observe high degrees of collaboration during the
  test exam, which is why we restrained from using test exam scores as a
  proxy for individual skill.}

The first row of table 3 shows our main variable of interest (
\emph{Self-selection} ) concerning the formation effect. We define the
difference in the estimate of \emph{Self-selection} between columns 1
and 3 as the composition effect. Table 3 reveals that in terms of group
performance, the formation effect is larger than the composition effect.
We estimate a formation effect of -4.2 percentage points, corresponding
to about .3 standard deviations of the project performance distribution.
Self-selected groups perform 4.2 percentage points worse on the projects
than randomly formed groups due to the formation effect. The composition
effect is -.9 percentage points in magnitude, indicating that
self-selected groups have a detrimental skill, gender, and home region
composition for productivity.\footnote{The composition effect is not
  statistically significant. By conducting a test, as suggested by Yan,
  Aseltine, and Harel (2013), to compare the estimators for
  self-selection in the two regressions with and without controls, we
  receive a p-value of .47.} From table 3, we cannot reject the null in
favor of hypothesis 1a or 1b.

As we see from columns 2 and 3, most of the composition effect stems
from a different skill composition in self-selected groups, e.g.,
self-selected groups are more skill homogeneous. A higher skill
homogeneity reduces the likelihood of having someone with a solid high
school GPA in one's group, which hurts group performance. The better the
best group member, measured by her high school GPA, the better the group
performance.\footnote{Keep in mind that good grades correspond to a low
  GPA.} Similarly, the individual with the second best high school GPA
affects group performance positively, but with half of the magnitude.
The difference in high school GPA between the individual with a group's
highest and lowest GPA explains little of the variation in group
performance. Note that groups with a large maximum within-group GPA
difference also have a higher probability of holding a member with a low
GPA, which presumably does not contribute to a better group performance.
Having a highly skilled individual in a group enhances group
performance. Such better group performance might occur through knowledge
spillovers, but more likely because the highly skilled individual solves
the most difficult tasks of a project. Suppose those skill spillovers
and the distribution of work are similar in self-selected and randomly
formed groups. In that case, skill composition explains about 15\% of
the performance loss in self-selected groups. We also ran a regression
using the exact specification that we pre-registered before collecting
the data in column 5. Using this specification, we find larger estimates
than in the specifications from columns 2 and 3, where we included
project fixed effects. However, because the two projects within a
semester are slightly different, we prefer to include these
project-fixed effects.

We conclude that in terms of productivity, the composition effect and
the formation effect lead to lower productivity in self-selected groups.
The formation effect in our setting is four times the size of the
composition effect. The composition effect implies that self-selected
groups have a productivity-harming group composition and/or interact
less efficiently and use fewer personal characteristics and skills. The
size and the direction of the estimate for the formation effect are
unexpected and contradict Kiessling, Radbruch, and Schaube (2022) and
Coricelli, Fehr, and Fellner (2004). We expect that allowing students to
choose group members motivates them to exert more effort. An alternative
explanation for our formation effect: We might overestimate the
formation effect by defining it as the remaining effect of group
formation after controlling for skill, gender, and home region
composition. Although we control group composition's most relevant
factors, we might miss elements of group composition uncorrelated with
skill, gender, and home region, differently distributed among
self-selected and random groups but crucial for group collaboration.

\subsection{Knowledge Acquisition}

Letting students form groups by themselves negatively affects group
performance, but what about individual learning? Does choosing your
teammates makes you learn more? In figure 4, we plot the knowledge
acquisition in self-selected and randomly formed groups separately.

In figure 4 ,we show that individuals from self-selected groups seem to
acquire slightly more knowledge during the group projects than those
from randomly formed groups.\footnote{Note that these two means are
  statistically not different from each other.} Although randomly formed
groups outperform self-selected groups on group projects, this does not
increase knowledge acquisition. Again, run similar regressions to those
in table 3 with this data, deviating in two points: First, in these
regressions, we observe individuals in one semester instead of groups in
a project. Therefore, in all five regressions in table 4, a student in a
semester denotes an observation. Second, as a consequence, we do not
include project fixed effects.

As in table 3 above, we present in the first line of table 4 the
estimate for the formation effect in column 3. Again, the difference
between the estimators of columns 1 and 3 shows the composition effect.
Table 4 reveals that the formation effect is larger than the composition
effect in absolute value, meaning self-selection has a net positive
effect on individual learning. The formation effect is positive in all
regressions, meaning self-selection helps individual learning. It is
economically meaningful with 5.5 percentage points (in column 4),
corresponding to .3 standard deviations of the knowledge acquisition
distribution. The difference between columns 1 and 3 shows the
composition effect, which is the difference between self-selected and
randomly formed groups due to their different compositions.\footnote{This
  difference is statistically not significant when conducting the test
  suggested in Yan, Aseltine, and Harel (2013), we receive a p-value of
  .13.} The composition effect is negative and 2.2 percentage points in
size, meaning self-selection hurts individual learning by creating less
diverse groups.

\begin{figure}
\centering
\includegraphics{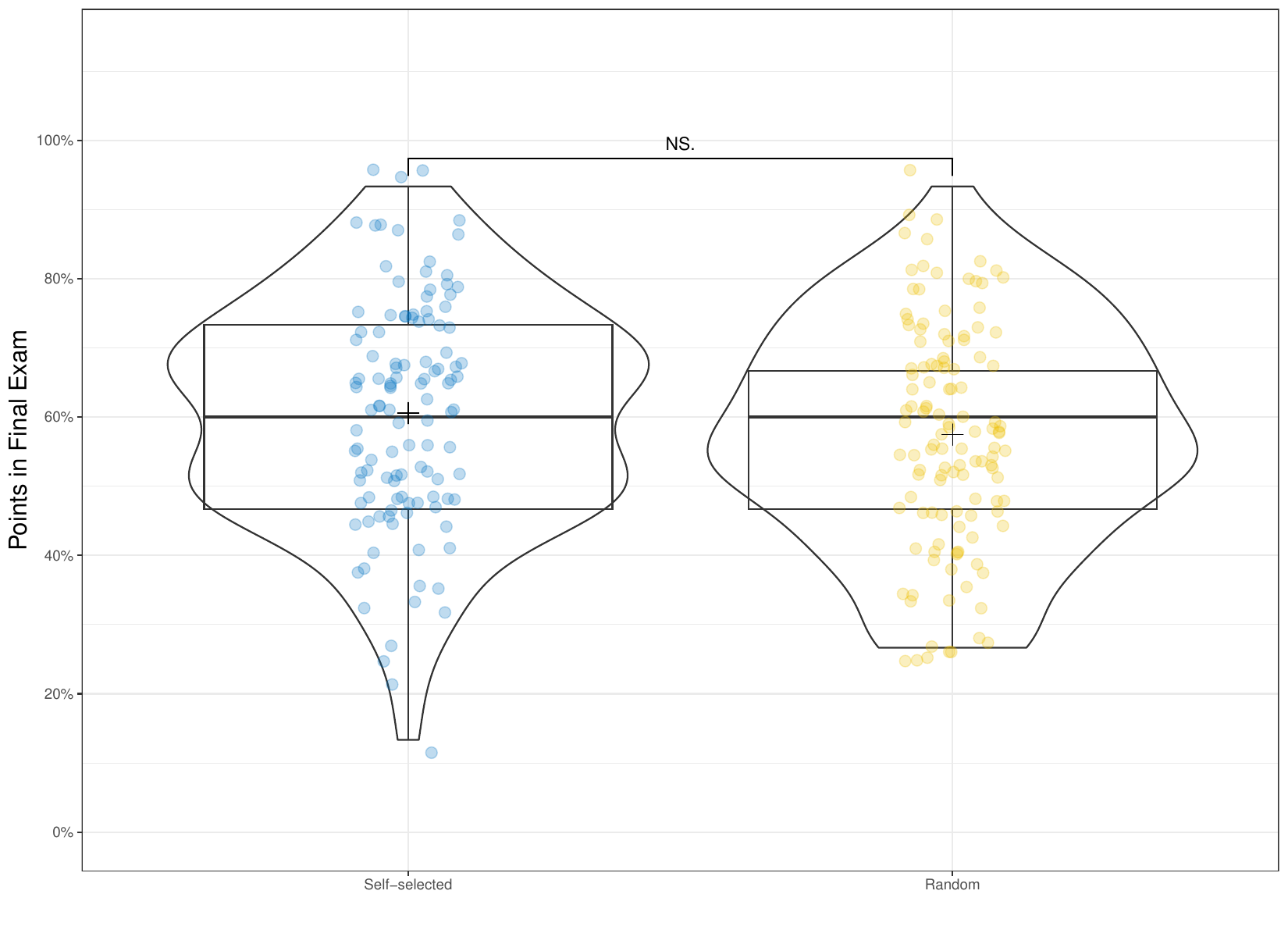}
\caption{Relationship between Knowledge Acquisition and Treatment}
\end{figure}

\vspace{-0.5cm}

\begin{singlespacing}

Notes: Each dot represents a student and a semester. We mark the mean with a cross. Further, we plot a kernel density and a boxplot. We measure knowledge acquisition by the number of points a student achieves in the final exam on project-related multiple-choice questions. There are 15 questions related to the projects in the final exam, each awarded with one point. For our analysis, we transform this knowledge acquisition measure to a scale from 0 - 100. N.S. stands for no statistically significant difference. 

\end{singlespacing}

We control for the skill level of individuals in a group with three
different measures: The high school GPA of the best student in the
group, the high school GPA of the second best student in the group, and
the difference in high school GPA between the best and the worst student
in the group. Having a highly skilled student in the group boosts
individual learning, probably because of knowledge spillovers. The other
two measures of skill composition (second-best GPA and Difference in GPA
) have no additional significant effect on individual learning, as shown
in column 2. We also find that gender and home region diversity matter
for individual learning. Students learn more when they work with
teammates of different gender and home regions than with teammates of
the same gender and home region, as shown in column 3 of table 4.

In column 4, we run our pre-registered regression on individual
knowledge acquisition from our pre-analysis plan. Column 4 shows that we
find larger estimates in the pre-registered specification than in the
specifications from columns 2 and 3. However, because our proxy for
ability (percentage points on the exam) is very noisy, as explained
above, our preferred specification is the one in column 3, where we use
student GPA and additionally control for other factors that affect group
composition. Following our analysis in table 4, we reject the null in
favor of hypothesis 2a but not in favor of 2b.

We conclude that the self-selection of groups has a complex impact on
individual learning. Due to the formation effect, individuals learn more
in self-selected groups. However, this positive effect is counteracted
by a less diverse group composition, which hinders individual learning.

\begin{table}[H] \centering 
  \caption{Effect of group-forming mechanism on final exam points on project questions} 
  \label{} 
\begin{tabular}{@{\extracolsep{5pt}}lcccc} 
\\[-1.8ex]\hline 
\hline \\[-1.8ex] 
 & \multicolumn{4}{c}{\textit{Dependent variable:}} \\ 
\cline{2-5} 
\\[-1.8ex] & \multicolumn{4}{c}{Percentage Points Final Exam Project Questions} \\ 
\\[-1.8ex] & (1) & (2) & (3) & (4)\\ 
\hline \\[-1.8ex] 
 Self-selection & 3.325 & 3.924$^{**}$ & 5.498$^{***}$ & 6.039$^{**}$ \\ 
  & (2.127) & (1.875) & (1.848) & (2.484) \\ 
  & & & & \\ 
 Best Group GPA &  & $-$10.471$^{***}$ & $-$13.836$^{***}$ &  \\ 
  &  & (3.698) & (3.274) &  \\ 
  & & & & \\ 
 Second Best Group GPA &  & $-$0.612 & 1.737 &  \\ 
  &  & (3.322) & (3.122) &  \\ 
  & & & & \\ 
 Max GPA Difference &  & $-$1.744 & $-$4.206$^{*}$ &  \\ 
  &  & (2.539) & (2.526) &  \\ 
  & & & & \\ 
 All Members Female &  &  & $-$9.051$^{**}$ &  \\ 
  &  &  & (3.974) &  \\ 
  & & & & \\ 
 One Member Male &  &  & $-$4.116 &  \\ 
  &  &  & (2.653) &  \\ 
  & & & & \\ 
 All Members Male &  &  & $-$1.621 &  \\ 
  &  &  & (2.150) &  \\ 
  & & & & \\ 
 Two Members Same Region &  &  & $-$3.869$^{**}$ &  \\ 
  &  &  & (1.836) &  \\ 
  & & & & \\ 
 Percentage Points Test Exam &  &  &  & 0.345$^{***}$ \\ 
  &  &  &  & (0.078) \\ 
  & & & & \\ 
 Self-selection:Summer Term &  &  &  & $-$7.108$^{*}$ \\ 
  &  &  &  & (3.820) \\ 
  & & & & \\ 
\hline \\[-1.8ex] 
Term FE & X & X & X & X \\ 
Observations & 241 & 241 & 241 & 239 \\ 
R$^{2}$ & 0.069 & 0.152 & 0.190 & 0.163 \\ 
Adjusted R$^{2}$ & 0.062 & 0.134 & 0.158 & 0.149 \\ 
Residual Std. Error & 15.713 & 15.097 & 14.880 & 15.005 \\ 
\hline 
\hline \\[-1.8ex] 
\multicolumn{5}{l}{$^{*}$p$<$0.1; $^{**}$p$<$0.05; $^{***}$p$<$0.01} \\ 
\multicolumn{5}{l}{\parbox[t]{12cm}{Notes: Standard errors reported in parentheses are clustered at the group level. In the regressions, a student in a semester define an observation. The dependent variable is the number of correct answers on the 15 project-related questions in the final exam, transformed to a 0 to 100 scale. Best Group GPA is a control variable for the best high school GPA in a group, and Second Best Group GPA for the second best high school GPA in a group. German high school GPAs range from 1 to 4, where 1 refers to the best grade and 4 to the worst. Max GPA Difference is a control variable for the difference in high school GPA between the member of the group with the best GPA and the one with the worst GPA. Two members same region is 1 if two members of a group graduated from high school in the same or neighboring counties, otherwise it is 0.}} \\ 
\end{tabular} 
\end{table}

We conjecture that self-selected groups interact differently with each
other and probably also distribute work differently than gender and home
region heterogeneous groups. This might positively affect group
performance but affect knowledge acquisition negatively. In the next
section, we provide evidence on how self-selected and randomly formed
groups differ in collaboration by analyzing individual contributions to
the group projects and knowledge acquisition.

\subsection{Satisfaction}

We have seen that self-selection affects group performance and
individual learning through adverse group composition but mostly through
the group-forming mechanism itself. We argue that a group-forming
mechanism only achieves high group performances and high levels of
knowledge gain in the long term if group members are content with their
group and the collaboration within the group. Therefore, we examine in
this section if choosing your teammates makes you happier.

\begin{figure}
\centering
\includegraphics{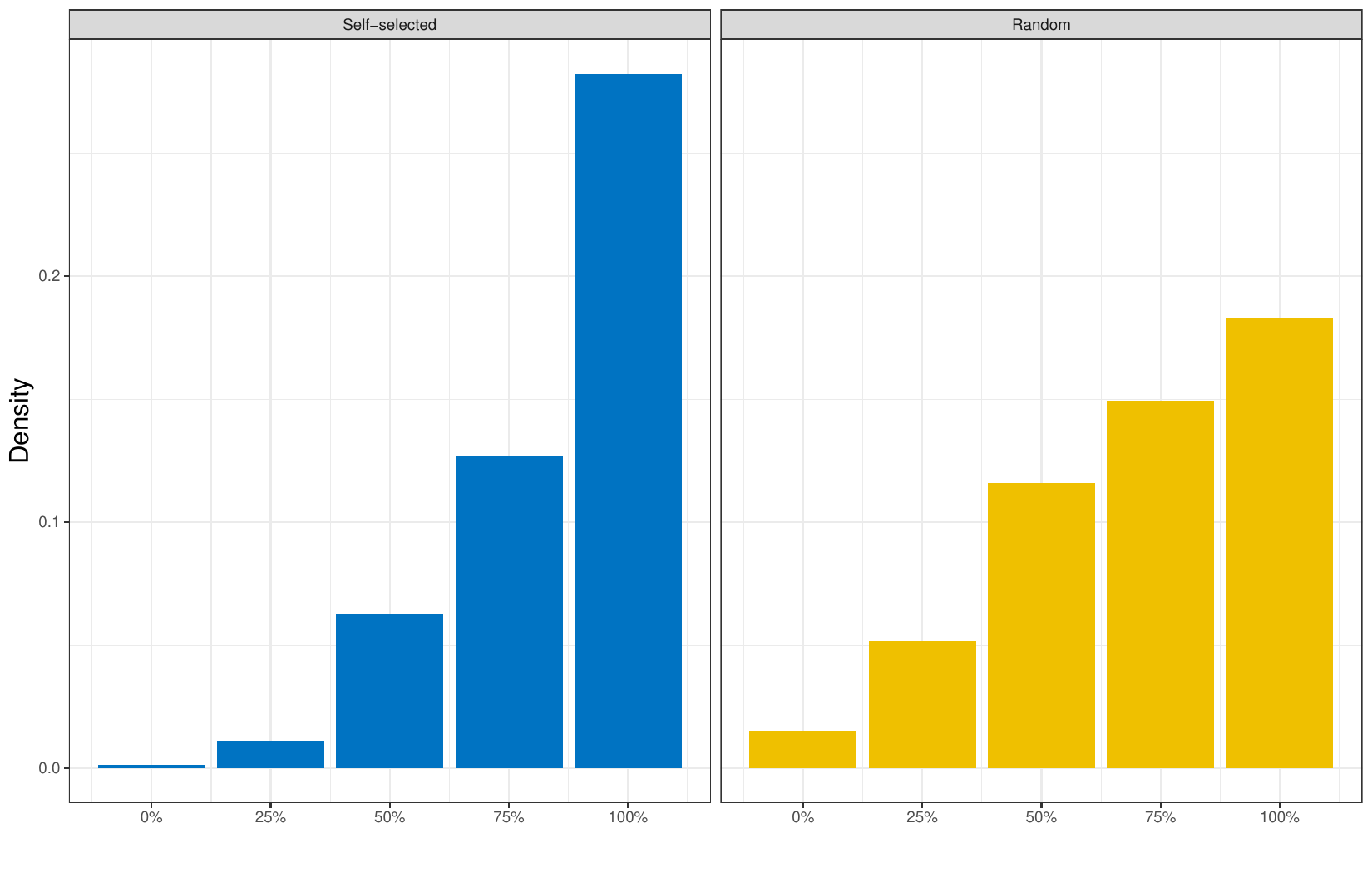}
\caption{Relationship between Satsifaction and Treatment}
\end{figure}

\vspace{-1cm}
\begin{singlespacing}

Notes: This figure shows the students' satisfaction with their team after each project. We elicit satisfaction in the surveys after each project on a linear 1 to 5 scale, where 5 refers to the highest satisfaction value. For our analysis, we transform this satisfaction measure to a scale from 0 - 100.

\end{singlespacing}

We look at student satisfaction with their groups from the surveys to
answer this question. We summarize our findings graphically in figure 5
and with OLS regressions in table 5. Figure 5 shows a shift towards
higher satisfaction levels for students in self-selected groups: Members
of self-selected groups report much higher satisfaction with the
teamwork than members of randomly formed groups. In table 5 we present
four different OLS regressions that estimate the effect of
self-selection on individual satisfaction. In these regressions, a
student in a semester denotes one observation. All regressions include
term fixed effects, and we cluster standard errors at the group level.

\begin{table}[H] \centering 
  \caption{Effect of group-forming mechanism on individual satisfaction} 
  \label{} 
\begin{tabular}{@{\extracolsep{-5pt}}lcccc} 
\\[-1.8ex]\hline 
\hline \\[-1.8ex] 
 & \multicolumn{4}{c}{\textit{Dependent variable:}} \\ 
\cline{2-5} 
\\[-1.8ex] & \multicolumn{4}{c}{\shortstack{Mean Satisfaction per Semester \\ in Percentage Points}} \\ 
\\[-1.8ex] & (1) & (2) & (3) & (4)\\ 
\hline \\[-1.8ex] 
 Self-selection & 12.429$^{***}$ & 13.382$^{***}$ & 13.182$^{***}$ & 8.628$^{**}$ \\ 
  & (2.920) & (2.778) & (2.674) & (4.033) \\ 
  & & & & \\ 
 Best Group GPA &  & $-$9.691$^{*}$ & $-$9.510 &  \\ 
  &  & (5.428) & (5.923) &  \\ 
  & & & & \\ 
 Second Best Group GPA &  & $-$1.299 & $-$1.462 &  \\ 
  &  & (4.274) & (4.861) &  \\ 
  & & & & \\ 
 Max GPA Difference &  & $-$0.114 & $-$0.433 &  \\ 
  &  & (4.758) & (4.962) &  \\ 
  & & & & \\ 
 All Members Female &  &  & $-$5.414 &  \\ 
  &  &  & (4.937) &  \\ 
  & & & & \\ 
 One Member Male &  &  & $-$1.097 &  \\ 
  &  &  & (4.142) &  \\ 
  & & & & \\ 
 All Members Male &  &  & $-$0.179 &  \\ 
  &  &  & (3.711) &  \\ 
  & & & & \\ 
 Two Members Same Region &  &  & 3.535 &  \\ 
  &  &  & (2.692) &  \\ 
  & & & & \\ 
 Percentage Points Test Exam &  &  &  & 0.215$^{**}$ \\ 
  &  &  &  & (0.099) \\ 
  & & & & \\ 
 Self-selection:Summer Term &  &  &  & 7.448 \\ 
  &  &  &  & (5.560) \\ 
  & & & & \\ 
\hline \\[-1.8ex] 
Term FE & X & X & X & X \\ 
Observations & 242 & 242 & 242 & 240 \\ 
R$^{2}$ & 0.102 & 0.164 & 0.176 & 0.128 \\ 
Adjusted R$^{2}$ & 0.094 & 0.147 & 0.145 & 0.113 \\ 
Residual Std. Error & 18.602 & 18.053 & 18.077 & 18.445 \\ 
\hline 
\hline \\[-1.8ex] 
\multicolumn{5}{l}{$^{*}$p$<$0.1; $^{**}$p$<$0.05; $^{***}$p$<$0.01} \\ 
\multicolumn{5}{l}{\parbox[t]{12cm}{Notes: Standard errors reported in parentheses are clustered at the group level. In the regressions, a student and a project define an observation. The dependent variable is the students' satisfaction with the team from the survey after each project, transformed to a 0 to 100 scale. Best Group GPA is a control variable for the best high school GPA in a group, and Second Best Group GPA for the second best high school GPA in a group. German high school GPAs range from 1 to 4, where 1 refers to the best grade and 4 to the worst. Max GPA Difference is a control variable for the difference in high school GPA between the member of the group with the best GPA and the one with the worst GPA. Two members same region is 1 if two members of a group graduated from high school in the same or neighboring counties, otherwise it is 0.}} \\ 
\end{tabular} 
\end{table}

In column 1 of table 5, we show the baseline regression that only
includes self-selection as an explanatory variable. In column 2, we add
skill composition variables. In column 3, we add gender and home region
composition variables, and in column 4, we run the pre-registered
regression from our pre-analysis plan.

Again, we show in the first line of table 5 the estimate for the
formation effect in column 3. The difference of the variable
\emph{Self-selection} between columns 1 and 3 shows the composition
effect. In terms of satisfaction, the formation effect dominates the
composition effect. Our estimate for the formation effect is 13.8
percentage points (in column 3), corresponding to roughly 2/3 standard
deviations of the satisfaction distribution. The composition effect is
negative, meaning self-selection hurts individual satisfaction by
creating less diverse and less optimal groups.\footnote{The composition
  effect statistically not significant with a p-value of .62 when
  conducting the test suggested in Yan, Aseltine, and Harel (2013).} The
formation effect is much larger than the composition effect in absolute
value, meaning that self-selection has a net positive effect on
individual satisfaction. In column 4, we run our pre-registered
regression on individual satisfaction from our pre-analysis plan. We
find smaller, but still sizeable, estimates in the pre-registered
specification than in the specifications from columns 2, 3, and 4.
Following our analysis in table 5, we reject the null in favor of
hypothesis 3a but not in favor of 3b.

Having a highly skilled student in the group ( \emph{Best Group GPA} )
boosts individual satisfaction, probably because of higher expectations
and confidence. We also find that gender and home region diversity do
not significantly affect individual satisfaction. Students are equally
happy or unhappy with their groups regardless of their teammates'
genders and home regions. Self-selection has a simple and robust impact
on individual satisfaction: It makes students happier with their groups.

\section{Mechanism}

In the results section, we show that self-selected groups have lower
group performance, acquire more knowledge, and are more satisfied with
their groups. These effects remain large when controlling for skill and
gender composition in groups. To examine the mechanisms behind these
effects, we analyze high school GPAs, individual contributions via
GitHub, and students' survey responses and relate them to group
performance, knowledge acquisition, and satisfaction. In this section,
we explore how groups allocate their work among their members depending
on how they formed groups and how this affects their group project
performance, knowledge acquisition, and satisfaction.

\subsection{Project performance and knowledge acquisition}

An unequal distribution of work may harm group performance and knowledge
acquisition. We, therefore, want to know how groups distribute work
within the group in randomly and self-selected groups. We use the
within-group standard deviation of the share of words committed to a
project on GitHub by each group member as a measure of work inequality.
A higher standard deviation means more unequal work distribution. We run
the pre-registered regression 3 from our empirical specification with
standard deviation of individual contributions as the dependent
variable. Each group in a project is one observation.

\begin{table}[H] \centering 
  \caption{Effect of group-forming mechanism on GitHub-Contributions} 
  \label{} 
\begin{tabular}{@{\extracolsep{5pt}}lcccc} 
\\[-1.8ex]\hline 
\hline \\[-1.8ex] 
 & \multicolumn{4}{c}{\textit{Dependent variable:}} \\ 
\cline{2-5} 
\\[-1.8ex] & \multicolumn{4}{c}{Within-Group SD of Share of Words} \\ 
\\[-1.8ex] & (1) & (2) & (3) & (4)\\ 
\hline \\[-1.8ex] 
 Self-selection & 0.160 & 0.644 & $-$1.371 & 3.916 \\ 
  & (2.971) & (3.118) & (3.370) & (4.104) \\ 
  & & & & \\ 
 Best Group GPA &  & 0.448 & 2.369 &  \\ 
  &  & (6.513) & (6.319) &  \\ 
  & & & & \\ 
 Second Best Group GPA &  & 3.640 & 0.716 &  \\ 
  &  & (5.699) & (5.425) &  \\ 
  & & & & \\ 
 Max GPA Difference &  & 2.195 & $-$0.039 &  \\ 
  &  & (4.201) & (4.101) &  \\ 
  & & & & \\ 
 Group Mean Test Exam &  &  &  & $-$0.125 \\ 
  &  &  &  & (0.167) \\ 
  & & & & \\ 
 Self-selection:Summer Term &  &  &  & $-$8.183 \\ 
  &  &  &  & (6.030) \\ 
  & & & & \\ 
\hline \\[-1.8ex] 
Gender Controls &  &  & X &  \\ 
Home Region Controls &  &  & X &  \\ 
Project FE & X & X & X &  \\ 
Term FE & X & X & X & X \\ 
Observations & 168 & 168 & 168 & 168 \\ 
R$^{2}$ & 0.019 & 0.038 & 0.127 & 0.039 \\ 
Adjusted R$^{2}$ & $-$0.005 & $-$0.004 & 0.065 & 0.015 \\ 
Residual Std. Error & 14.713 & 14.702 & 14.188 & 14.562 \\ 
\hline 
\hline \\[-1.8ex] 
\multicolumn{5}{l}{$^{*}$p$<$0.1; $^{**}$p$<$0.05; $^{***}$p$<$0.01} \\ 
\multicolumn{5}{l}{\parbox[t]{11cm}{ Notes: Standard errors reported in parentheses are clustered at the group level. In the regressions, a group and a project define an observation. For the dependent variable we calculate the share of words from the number of words of code and text a student has contributed via GitHub to the total amount of words of code and text of her or his group in a project. We then calculate for each group and project the standard deviation of the shares of words of the group members, which is our dependent variable in these regressions. Best Group GPA is a control variable for the best high school GPA in a group, and Second Best Group GPA for the second best high school GPA in a group. German high school GPAs range from 1 to 4, where 1 refers to the best grade and 4 to the worst. Max GPA Difference is a control variable for the difference in high school GPA between the member of the group with the best GPA and the one with the worst GPA. Two members same region is 1 if two members of a group graduated from high school in the same or neighboring counties, otherwise it is 0.}} \\ 
\end{tabular} 
\end{table}

\vspace{1.0cm}

We find no evidence in table 6 that self-selected or randomly formed
groups distribute the work within their groups more unequally. We also
find no evidence that the group skill composition affects the work
distribution, and therefore, we cannot reject the null in favor of
hypothesis 4 or 4a. For robustness, we measure inequality in work
distribution per group with the Herfindahl-index but find no significant
differences between randomly and self-selected groups.\footnote{We
  present the results in Appendix D.}

Albeit groups distribute work not more or less inequally in one of the
two treatments, they might distribute work to group members with
particular characteristics (e.g.~the more skilled group members).
Moreover, social ties and communication in self-selected groups might
change the impact of high-skilled individuals on group and individual
outcomes.

From our main results we know that highly skilled students in groups
enhance group performance significantly. This enhanced group performance
could result from the fact that highly skilled students transfer their
knowledge to other group members to achieve a higher performance as a
team. Alternatively, the highly skilled group member could solve more
tasks overall or more challenging tasks in a project by himself. To
investigate this channel, we look at the individual contributions of
each student to their GitHub projects. We compare self-selected and
randomly formed groups to see if the best student in the group has a
different role depending on how the group was formed. In table 7 we
examine the individual contributions and outcomes of students in
self-selected and randomly formed groups. We use two dummy variables to
indicate whether a student has the best or the second-best high school
GPA in their group. We look at the share of words committed by each
student to their GitHub project (columns 1 to 4) and their knowledge
acquisition, measured by the score in the final exam on project-related
questions (columns 5 to 8). We control for the skill distribution, the
home region, and the gender composition of each group.

\begin{table}[H] \centering    \caption{Effect of group-forming mechanism on final exam points on project questions for different GPAs}    \label{}  \resizebox{0.98\textwidth}{!}{\begin{tabular}{@{\extracolsep{5pt}}lcccccccc}  \\[-1.8ex]\hline  \hline \\[-1.8ex]   & \multicolumn{8}{c}{\textit{Dependent variable:}} \\  \cline{2-9}  \\[-1.8ex] & \multicolumn{4}{c}{Share of GitHub Contributions} & \multicolumn{4}{c}{Percentage Points in Final Exam on Project Questions} \\   & \multicolumn{2}{c}{(Self-select)} & \multicolumn{2}{c}{(Random)} & \multicolumn{2}{c}{(Self-select)} & \multicolumn{2}{c}{(Random)} \\  \hline \\[-1.8ex]   Own is Best Group GPA & 13.213$^{**}$ & 13.709$^{**}$ & 24.711$^{***}$ & 24.717$^{***}$ & 5.526 & 6.114 & 10.640$^{***}$ & 10.759$^{***}$ \\    & (5.886) & (5.989) & (5.772) & (5.862) & (3.708) & (3.684) & (3.334) & (3.448) \\    & & & & & & & & \\   Own is Second Best Group GPA & 4.001 & 4.385 & 9.556$^{**}$ & 9.600$^{**}$ & $-$0.728 & 0.024 & 4.767 & 5.082 \\    & (6.083) & (6.218) & (3.990) & (4.077) & (3.638) & (3.698) & (3.506) & (3.617) \\    & & & & & & & & \\   Best Group GPA &  & 5.173 &  & $-$0.703 &  & $-$17.879$^{***}$ &  & $-$8.872$^{**}$ \\    &  & (5.429) &  & (1.349) &  & (5.888) &  & (3.308) \\    & & & & & & & & \\   Second Best Group GPA &  & $-$5.712 &  & 0.709 &  & 0.371 &  & 2.645 \\    &  & (4.655) &  & (1.470) &  & (5.215) &  & (3.543) \\    & & & & & & & & \\   Max GPA Difference &  & $-$1.113 &  & 0.867 &  & $-$7.843$^{**}$ &  & 1.000 \\    &  & (2.898) &  & (1.050) &  & (3.491) &  & (2.759) \\    & & & & & & & & \\  \hline \\[-1.8ex]  Home Region Controls &  & X &  & X &  & X &  & X \\  Gender Controls &  & X &  & X &  & X &  & X \\  Term FE & X & X & X & X & X & X & X & X \\  Project FE & X & X & X & X &  &  &  &  \\  Observations & 242 & 242 & 248 & 248 & 117 & 117 & 124 & 124 \\  R$^{2}$ & 0.048 & 0.059 & 0.199 & 0.200 & 0.043 & 0.272 & 0.214 & 0.296 \\  Adjusted R$^{2}$ & 0.028 & 0.009 & 0.183 & 0.159 & 0.017 & 0.203 & 0.194 & 0.234 \\  Residual Std. Error & 24.639 & 24.872 & 20.611 & 20.907 & 16.156 & 14.547 & 14.415 & 14.054 \\  \hline  \hline \\[-1.8ex]  \multicolumn{9}{l}{$^{*}$p$<$0.1; $^{**}$p$<$0.05; $^{***}$p$<$0.01} \\  \multicolumn{9}{l}{\parbox[t]{22cm}{ Notes: Standard errors reported in parentheses are clustered at the group level. In the regressions, a student in a semester define an observation. In colums 1 to 4, the dependent variable is the share of words contributed by a student to a project, and in columns 5 to 8, the number of correct answers on the 15 project-related questions in the final exam, transformed to a 0 to 100 scale. Own is Best Group GPA and Own is Second Best Group GPA indicate whether a student has the best or the second best high school GPA in their group. Best Group GPA is a control variable for the best high school GPA in a group, and Second Best Group GPA for the second best high school GPA in a group. German high school GPAs range from 1 to 4, where 1 refers to the best grade and 4 to the worst. Max GPA Difference is a control variable for the difference in high school GPA between the member of the group with the best GPA and the one with the worst GPA. Two members same region is 1 if two members of a group graduated from high school in the same or neighboring counties, otherwise it is 0. Individual With Best GPA in Group is 1 if the corresponding student has the best high school GPA in the group, otherwise it is 0.}} \\  \end{tabular}}  \end{table}

We show, in table 7, that the best student in a group contributes more
than the second-best and the worst student in both self-selected and
randomly formed groups. This is consistent with our findings above in
table 6. The most skilled group member contributes roughly 15 percentage
points more than the lowest skilled group member in self-selected
groups, whereas the most skilled group member contributes 22 percentage
points more in randomly formed groups.\footnote{In Appendix E we show
  that the more skilled members of a group contribute not significantly
  more to the challenging tasks of a project.}

We also find that the most skilled student in a self-selected group
acquires not significantly more knowledge than their group members, even
if we control for the skill distribution, the home region, and the
gender composition in the group. This suggests that there are skill
spillovers in self-selected groups. We do not find spillovers to the
same extent in randomly formed groups, where the best student performs
significantly better than the others.

\begin{table}[H] \centering 
  \caption{Effect of group-forming mechanism and contribution in project on knowledge aquisition} 
  \label{} 
\begin{tabular}{@{\extracolsep{5pt}}lcc} 
\\[-1.8ex]\hline 
\hline \\[-1.8ex] 
 & \multicolumn{2}{c}{\textit{Dependent variable:}} \\ 
\cline{2-3} 
\\[-1.8ex] & \multicolumn{2}{c}{\shortstack{Percentage Points in Final Exam \\ on Project Questions}} \\ 
\hline \\[-1.8ex] 
 Self-selection & 8.775$^{***}$ & 11.155$^{***}$ \\ 
  & (3.242) & (3.276) \\ 
  & & \\ 
 Share of Github Contributions in Group & 0.277$^{***}$ & 0.275$^{***}$ \\ 
  & (0.059) & (0.060) \\ 
  & & \\ 
 Self-selection x Share of Github Contributions & $-$0.159$^{**}$ & $-$0.165$^{**}$ \\ 
  & (0.077) & (0.077) \\ 
  & & \\ 
 Best Group GPA &  & $-$13.685$^{***}$ \\ 
  &  & (3.319) \\ 
  & & \\ 
 Second Best Group GPA &  & 1.801 \\ 
  &  & (3.143) \\ 
  & & \\ 
 Max GPA Difference &  & $-$4.030 \\ 
  &  & (2.493) \\ 
  & & \\ 
\hline \\[-1.8ex] 
Home Region Controls &  & X \\ 
Gender Controls &  & X \\ 
Term FE & X & X \\ 
Observations & 241 & 241 \\ 
R$^{2}$ & 0.152 & 0.270 \\ 
Adjusted R$^{2}$ & 0.137 & 0.235 \\ 
Residual Std. Error & 15.066 & 14.189 \\ 
\hline 
\hline \\[-1.8ex] 
\multicolumn{3}{l}{$^{*}$p$<$0.1; $^{**}$p$<$0.05; $^{***}$p$<$0.01} \\ 
\multicolumn{3}{l}{\parbox[t]{11cm}{Notes: Standard errors reported in parentheses are clustered at the group level. In the regressions, a student in a semester define an observation. The dependent variable is the number of correct answers on the 15 project-related questions in the final exam, transformed to a 0 to 100 scale. Share of Github Contributions in Group ranges from 0 to 100 and indicates which share of a groups words of code and text committed to all of the groups projects throughout a semester is provided by the corresponding student.}} \\ 
\end{tabular} 
\end{table}

Table 7 reveals, that the most skilled members of a group contribute
more to the overall project and acquire more knowledge, but only in
randomly formed groups. In self-selected groups, both effects are less
pronounced. In the next step, we examine whether contributing to a
project is associated with higher knowledge acquisition or if the highly
skilled members acquire knowledge well due to other reasons than
contributing substantially to the projects. We, therefore, regress in
table 8 knowledge acquisition on students' contributions to the projects
of a group in a semester via GitHub and interact it with a group
formation dummy variable. We show that contributing one percentage point
more to the projects throughout a semester increases knowledge
acquisition in self-selected groups by 0.11 percentage points on average
and in randomly formed groups by 0.28. We conclude that in order to
acquire knowledge, it is significantly more important for students from
randomly formed groups to contribute to the projects than those from
self-selected groups.\footnote{In Appendix F, we show on the
  project-task level that contributing to a task of a project is
  associated with a higher probability of answering exam questions
  correctly, that relate to that question. This effect is less
  pronounced for self-selected groups.}

So why do self-selected groups perform worse on the group projects, but
individuals acquire more knowledge from the projects? Based on our
findings in this section, we come up with the following explanation: In
randomly formed groups, the most skilled student of a group contributes
significantly more to the projects than in self-selected groups. Because
in self-selected groups, students with low skills solve more project
tasks, they perform worse on the group projects than randomly formed
groups. However, we also see that in self-selected groups, the
low-skilled members and those who contribute little to the projects
acquire almost as much knowledge as the rest of the group. In contrast,
in randomly formed groups, students with low prior skills and those who
contribute little to the projects acquire only little knowledge from the
projects.\footnote{Remember, the skill level rank of a student within a
  group and contribution to a group project is negatively correlated.
  This means the most skilled members of a group do most of the work.}
We argue, that these differences between randomly and self-selected
groups occur because of better communication and discussion in
self-selected groups nourished by stronger social ties and higher group
identity. This then results in higher knowledge acquisition for the
entire group.

So far we find, that self-selected groups distribute work substantially
differently than randomly formed groups. In the tables above we show how
the different distribution of work in the two treatments explains the
formation effect. In table 9 we examine whether distributing work to
more or less skilled individuals works as a channel for explaining the
composition effect. Thus, we add to the regressions from table 3 and 4
control variables for the share of GitHub contributions by the members
with the highest and lowest skill levels in a group. We show in table 9
that the composition effect almost diminishes for group project
performance but remains nearly unchanged for knowledge
acquisition.\footnote{Note that in table 9 the estimator for the effect
  of the contribution of the most skilled group member on project
  performance and the the estimator for the effect of the contribution
  of the least skilled group member on knowledge acquisition are
  negative. We argue that the effects are highly heterogeneous and that
  having high contributions of these students is correlated with a
  dysfunctional group.} We conclude that how groups distribute work
across skill levels explains the majority of the composition effect in
group project performance. This means that the different group
composition in self-selected groups affects project performance mostly
because groups distribute work differently according to the skill levels
of their members. The composition effect in knowledge acquisition seems
to work differently: The effect of a different group composition in
self-selected groups on knowledge acquisition occurs not because groups
distribute work differently in self-selected groups. The skill level of
an individual's group members seems to directly impact the knowledge
acquisition of that individual. However, remember that we find above a
much larger formation effect which affects, through a different work
distribution in self-selected and randomly formed groups, group
performance and knowledge acquisition.

\begin{table}[H] \centering 
  \caption{Effect of group-forming mechanism on Group Performance and Knowledge Acquisition with Contribution controls} 
  \label{} 
\begin{tabular}{@{\extracolsep{5pt}}lcccc} 
\\[-1.8ex]\hline 
\hline \\[-1.8ex] 
 & \multicolumn{4}{c}{\textit{Dependent variable:}} \\ 
\cline{2-5} 
\\[-1.8ex] & \multicolumn{2}{c}{Project Percentage Points} & \multicolumn{2}{c}{\shortstack{Percentage Points in Final Exam \\ on Project Questions}} \\ 
\\[-1.8ex] & (1) & (2) & (3) & (4)\\ 
\hline \\[-1.8ex] 
 Self-selection & $-$5.134$^{**}$ & $-$4.910$^{**}$ & 3.325 & 5.279$^{***}$ \\ 
  & (2.418) & (2.307) & (2.127) & (1.804) \\ 
  & & & & \\ 
 Contribution of Best &  & $-$14.333$^{***}$ &  & $-$5.714 \\ 
  &  & (4.846) &  & (3.986) \\ 
  & & & & \\ 
 Contribution of Third-Best &  & 0.823 &  & $-$13.851$^{***}$ \\ 
  &  & (6.017) &  & (3.836) \\ 
  & & & & \\ 
\hline \\[-1.8ex] 
Skill Controls &  & X &  & X \\ 
Home Region Controls &  & X &  & X \\ 
Gender Controls &  & X &  & X \\ 
Project FE & X & X &  &  \\ 
Term FE & X & X & X & X \\ 
Observations & 172 & 172 & 241 & 241 \\ 
R$^{2}$ & 0.075 & 0.228 & 0.069 & 0.212 \\ 
Adjusted R$^{2}$ & 0.052 & 0.165 & 0.062 & 0.174 \\ 
Residual Std. Error & 13.896 & 13.045 & 15.713 & 14.743 \\ 
\hline 
\hline \\[-1.8ex] 
\multicolumn{5}{l}{$^{*}$p$<$0.1; $^{**}$p$<$0.05; $^{***}$p$<$0.01} \\ 
\multicolumn{5}{l}{\parbox[t]{14cm}{ Notes: Standard errors reported in parentheses are clustered at the group level. In columns 1 and 2, a group and a project define an observation and the dependent variable is the performance on the group projects on a scale from 0 to 100. In columns 3 and 4, a student and a semester define an observation and the dependent variable is the performance on the 15 questions related to the projects in the final exam on a scale from 0 to 100.  Contribution of Best and Contribution of Third-best indicate which share of the GitHub Commits was contributed by the group member with the highest high school GPA and which share by the group member with the lowest high school GPA in a group. }} \\ 
\end{tabular} 
\end{table}

\subsection{Satisfaction}

Self-selected groups are more satisfied because of a formation effect.
We come up with two possible explanations: First, individuals perceive
the extent of their own contribution in self-selected groups more
precisely and do not overestimate their contribution. Second,
individuals internalize the external effects of their contributions in
self-selected groups and are less dissatisfied with a higher perceived
contribution in self-selected groups than in randomly formed groups.
Figure 6 shows the relationship between perceived and actual
contributions to the project on GitHub on the left and between perceived
contribution and satisfaction on the right. We see that both
self-selected and randomly formed groups report their contributions
accurately. However, those who contribute and those who perceive to
contribute more than average are much happier in self-selected groups
than in randomly formed groups.

\begin{figure}[H]
\includegraphics[width=1\linewidth]{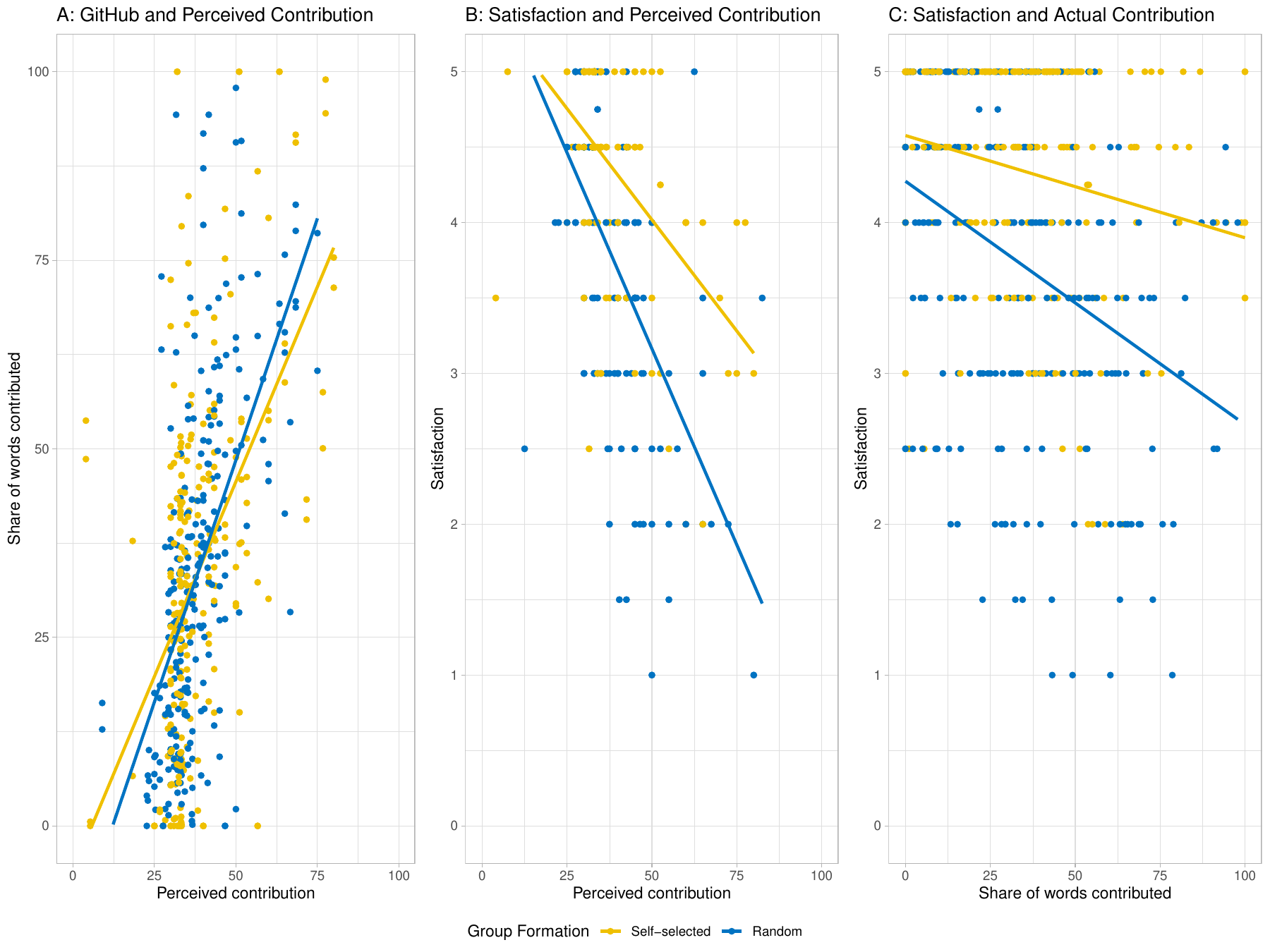} \caption{GitHub Contribution and Perceived Contribution in Semester}\label{fig:unnamed-chunk-20}
\end{figure}

\vspace{-1cm}

\begin{singlespacing}
Notes: We calculate the share of words contributed from the number of words a student contributes via GitHub to the total amount of words of their group in a project. We here do not distinguish between words of code and words of text contributed. We elicit students' satisfaction with the team in the survey after each project on a scale from 1 to 5, where 5 refers to high satisfaction levels. In the same surveys, we ask students to indicate on a scale from 0 to 100 how much they think they contributed to the group project. Each observation corresponds to an individual and a project.

\end{singlespacing}

In table 10, we regress satisfaction on students' contributions to the
projects of a group in a semester via GitHub. We show that contributing
one percentage point more to the projects throughout a semester
decreases satisfaction in self-selected groups by 0.148 percentage
points and in randomly formed groups by 0.369 percentage points. We
conjecture that students in self-selected groups are more satisfied when
contributing a large share of the overall project because they
internalize the benefits of their contribution and enjoy doing more work
if they can pick their group members.

\begin{table}[H] \centering 
  \caption{Effect of group-forming mechanism and contribution in project on satisfaction} 
  \label{} 
\begin{tabular}{@{\extracolsep{5pt}}lcc} 
\\[-1.8ex]\hline 
\hline \\[-1.8ex] 
 & \multicolumn{2}{c}{\textit{Dependent variable:}} \\ 
\cline{2-3} 
\\[-1.8ex] & \multicolumn{2}{c}{Satisfaction} \\ 
\hline \\[-1.8ex] 
 Self-selection & 4.896 & 5.585 \\ 
  & (4.267) & (4.270) \\ 
  & & \\ 
 Share of Github Contributions in Group & $-$0.369$^{***}$ & $-$0.373$^{***}$ \\ 
  & (0.091) & (0.094) \\ 
  & & \\ 
 Self-selection x Share of Github Contributions & 0.221$^{**}$ & 0.223$^{**}$ \\ 
  & (0.107) & (0.112) \\ 
  & & \\ 
 Best Group GPA &  & $-$9.704 \\ 
  &  & (5.919) \\ 
  & & \\ 
 Second Best Group GPA &  & $-$1.529 \\ 
  &  & (4.840) \\ 
  & & \\ 
 Max GPA Difference &  & $-$0.688 \\ 
  &  & (4.931) \\ 
  & & \\ 
\hline \\[-1.8ex] 
Home Region Controls &  & X \\ 
Gender Controls &  & X \\ 
Term FE & X & X \\ 
Observations & 242 & 242 \\ 
R$^{2}$ & 0.201 & 0.277 \\ 
Adjusted R$^{2}$ & 0.187 & 0.242 \\ 
Residual Std. Error & 17.622 & 17.010 \\ 
\hline 
\hline \\[-1.8ex] 
\multicolumn{3}{l}{$^{*}$p$<$0.1; $^{**}$p$<$0.05; $^{***}$p$<$0.01} \\ 
\multicolumn{3}{l}{\parbox[t]{12cm}{ Notes: Standard errors reported in parentheses are clustered at the group level. In the regressions, a student in a semester define an observation. The dependent variable is the average satisfaction level of a student in a semester from the surveys and ranges from 0 to 100. Share of Github Contributions in Group ranges from 0 to 100 and indicates which share of a groups words of code and text committed to all of the groups projects throughout a semester is provided by the corresponding student.}} \\ 
\end{tabular} 
\end{table}

\section{Robustness}

The regressions above on group performance analyze the effect of letting
groups form themselves on the grades of the two projects in each
semester and cohort. However, students might smooth their individual
contributions over a semester, not a project. This could affect the
individual effort a student assigns a project and the effort the whole
group puts into a project. In Appendix G, we run the regressions from
table 3 again and show that results remain unchanged if we aggregate
observations on semester and group level. In the regression table in
Appendix G each observation represents a group in a semester, whereas in
table 3 one observation denotes a group in a project.

How well a group performs might also depend on the social preferences
and social skills of its members.\footnote{See, e.g. Weidmann and Deming
  (2021)} We add dummy variables to our main regression from table 3 for
whether there is a team player in a group, whether there is someone with
altruistic social preferences in a group, and whether there is someone
with conditional cooperative social preferences in a group. We obtain
these measures from an online lab experiment we conduct for each cohort
at the end of the summer term. \footnote{We pay students 6 EUR to
  participate in the online experiment. The online experiment has three
  parts: an incentivized dictator game, a sequential prisoner's dilemma,
  and a shortened version of the modified Reading the Mind in the Eyes
  Test (RMET) to measure social preferences and cognitive empathy. We
  use the dictator game to control for altruistic behavior among the
  participants. In the dictator game, we ask participants to split 10
  EUR between themselves and a charity of their choice. Each participant
  is a dictator and there are no receivers. We use the sequential
  prisoner's dilemma to classify participants into three prosocial
  types: altruists, conditional cooperators, and selfish. Our version of
  the prisoner's dilemma is the same as Esteves-Sorenson (2018): First,
  we show each participant how payoffs are determined in a prisoner's
  dilemma. Then, we present them with a prisoner's dilemma with
  potential payoffs between 0.00 and 5.63 EUR. In the first round, all
  participants are first movers and in the second round, all
  participants are second movers and decide how they would react to each
  of the two first mover choices. Then, we match students randomly and
  implement their strategies. We use the RMET to assess emotional
  intelligence, which is strongly related to team efficiency. In the
  modified RMET by Baron-Cohen et al. (2001), participants have to read
  emotions by looking at pairs of eyes. For each pair of eyes, the
  participant has four options and only one is correct. We use the
  shortened version from Weidmann and Deming (2021) with 26 pairs of
  eyes. We also provide students with synonyms for unclear expressions.
  After the dictator game, the sequential prisoner's dilemma, and the
  RMET, we randomly choose one of the three parts to pay each
  participant. Fehr and Schmidt (1999) showed that subjects do not
  behave differently with probabilistic payments compared to a certain
  amount of money. Note that the RMET is not incentivized, and
  therefore, participants receive no extra payment if this part of the
  experiment is randomly chosen for payoff. At the end of the online
  experiment, students have to fill in a short survey to provide us with
  the necessary information to pay participants and match data from the
  field experiment.} In defining team players we follow Weidmann and
Deming (2021): A team player performs one standard deviation or more
above the mean in the RMET.\footnote{Note that RMET and high school GPA
  are not correlated, so we control for two covariates: Academic skill
  and cognitive empathy} We show in Appendix H, that adding those
additional covariates does not change our estimates from table 3. We do
not include these variables in our main regression results because
roughly 1/3 of the students did not participate in the experiment, and
we assume that these students are systematically less involved with the
course and have social preferences, deviating from the rest of the
sample.

We change the projects each semester and cohort. As a result, we have
for each of those projects only observations for one kind of group
formation mechanism. We are aware of the problem that different levels
of difficulty of the projects could drive our results on the effect of
group formation on project points and individual perception of the
course. We address this problem in two ways: First, even though we
change the projects, we keep them similar in the way we ask questions
and the skills required to work on them. Second, we hire a student
assistant, skilled in data analysis, to evaluate and compare the
difficulty levels of Project 1, 2, 3, and 4 in the two cohorts. We do
not inform the student assistant about the experiment. Table 11 shows
how the student assistant rates the difficulty levels of the projects
from cohort 2021 compared to the projects from cohort 2020. ``+'' means,
that the project was more difficult in cohort 2021 and ``-'' that it was
easier in cohort 2021. The evaluation scale is linear and ranges from
``- - -'' to ``+ + +''. Table 11 shows that taken both projects of a
semester together, the projects of winter term 2020 have the same level
of difficulty as those of winter term 2021, and those of summer term
2020 have the same level of difficulty as those of summer term 2021.

\renewcommand{\arraystretch}{2}
\begin{table}

\caption{\label{tab:unnamed-chunk-22}Comparison of Project Difficulty Levels}
\centering
\begin{tabu} to \linewidth {>{\raggedright}X>{\raggedright}X>{\raggedright}X>{\raggedright}X>{\raggedright}X}
\toprule
  & Project 1 & Project 2 & Project 3 & Project 4\\
\midrule
Cohort 2021 compared to 2020 & + + & - - & - - & +\\
\bottomrule
\end{tabu}
\end{table}
\renewcommand{\arraystretch}{1}

\section{Conclusion}

We conduct a natural field experiment in the classroom, where students
work in groups of three on cognitively challenging projects. We
investigate how the composition and formation effect drive differences
in the project performance of groups, knowledge gained by individual
group members, and individual satisfaction between self-selected and
randomly formed groups. We find that the formation effect dominates the
composition effect in all three dimensions: The formation effect is four
times the size of the composition effect in terms of group performance,
three times the size in terms of knowledge acquisition, and eleven times
the size in terms of satisfaction.

The composition effect is negative for self-selected groups in all three
observed dimensions. We conclude that self-selected groups have a
productivity, knowledge acquisition, and satisfaction-harming group
composition. This is the case because the elevated skill homogeneity in
self-selected groups harms productivity and satisfaction, and together
with the increased homogeneity in terms of gender and home region in
self-selected groups, it also harms knowledge acquisition. In contrast,
we associate the formation effect with lower group productivity but
higher knowledge acquisition and satisfaction. We explain these adverse
effects with differences in how groups distribute work across their
members: Randomly formed groups distribute more work to the highly
skilled group members, for whom the marginal effect of acquiring
knowledge from contributing to the projects is smaller than for less
skilled individuals. This explains why self-selected groups acquire more
knowledge during the group projects, albeit performing worse on the
projects than randomly formed groups. We further show that self-selected
groups are more satisfied because high-contributors are less unsatisfied
with contributing a lot when they can choose their group members.

If we take the formation and composition effect together, we find that
self-selected groups perform 5.1 percentage points worse in their
projects than randomly formed groups. This effect is statistically
significant and economically meaningful as it composes 1/3 of a standard
deviation in the project performance distribution.\footnote{See for
  instance Boss et al. (2021) with and effect size of .19 SD, Fenoll and
  Zaccagni (2022) with 0.38 SD, Fischer, Rilke, and Yurtoglu (2023) with
  0.52 SD, and Kiessling, Radbruch, and Schaube (2022) with 0.15 SD.}
Moreover, members of self-selected groups acquired 3.3 percentage points
more knowledge and are 12.5 percentage points more satisfied than those
in randomly formed groups. Our group performance results align with
Fischer, Rilke, and Yurtoglu (2023) on group presentations and
contradict Fenoll and Zaccagni (2022) and Kiessling, Radbruch, and
Schaube (2022). Compared to our setting, the task in Kiessling,
Radbruch, and Schaube (2022) is non-collaborative, and effort does not
affect the public good.

Our results implicate that self-selection on cooperative high cognition
tasks with exchangeable group members is performance-harming. However,
self-selection comes at a higher satisfaction and a lower workload of
highly skilled individuals compared to randomly formed groups. We
advocate for teaching settings in which knowledge acquisition is
essential to give individuals some choice in the group formation
process. Moreover, in settings where individual well-being is of major
interest, we recommend letting groups self-select.

\clearpage
\newpage

\section{References}

\hypertarget{refs}{}
\begin{CSLReferences}{1}{0}
\leavevmode\vadjust pre{\hypertarget{ref-aiPuttingTeamsGig2023}{}}%
Ai, Wei, Yan Chen, Qiaozhu Mei, Jieping Ye, and Lingyu Zhang. 2023.
{``Putting {Teams} into the {Gig Economy}: {A Field Experiment} at a
{Ride-sharing Platform}.''} \emph{Management Science}, 30.

\leavevmode\vadjust pre{\hypertarget{ref-bandieraSocialPreferencesResponse2005}{}}%
Bandiera, Oriana, Iwan Barankay, and Imran Rasul. 2005. {``Social
{Preferences} and the {Response} to {Incentives}: {Evidence} from
{Personnel Data}*.''} \emph{The Quarterly Journal of Economics} 120 (3):
917--62.

\leavevmode\vadjust pre{\hypertarget{ref-bandieraTeamIncentivesEvidence2013}{}}%
---------. 2013. {``Team {Incentives}: {Evidence} from a {Firm Level
Experiment}.''} \emph{Journal of the European Economic Association} 11
(5): 1079--1114.

\leavevmode\vadjust pre{\hypertarget{ref-baron-cohenReadingMindEyes2001}{}}%
Baron-Cohen, Simon, Sally Wheelwright, Jacqueline Hill, Yogini Raste,
and Ian Plumb. 2001. {``The {`{Reading} the {Mind} in the {Eyes}'} {Test
Revised Version}: {A Study} with {Normal Adults}, and {Adults} with
{Asperger Syndrome} or {High-functioning Autism}.''} \emph{Journal of
Child Psychology and Psychiatry} 42 (2): 241--51.

\leavevmode\vadjust pre{\hypertarget{ref-bossOrganizingEntrepreneurialTeams2021}{}}%
Boss, Viktoria, Linus Dahlander, Christoph Ihl, and Rajshri Jayaraman.
2021. {``Organizing {Entrepreneurial Teams}: {A Field Experiment} on
{Autonomy} over {Choosing Teams} and {Ideas}.''} \emph{Organization
Science}, November.

\leavevmode\vadjust pre{\hypertarget{ref-carrellNaturalVariationOptimal2013}{}}%
Carrell, Scott E., Bruce I. Sacerdote, and James E. West. 2013. {``From
{Natural Variation} to {Optimal Policy}? {The Importance} of {Endogenous
Peer Group Formation}.''} \emph{Econometrica} 81 (3): 855--82.

\leavevmode\vadjust pre{\hypertarget{ref-charroinPeerEffectsSelfselection2022}{}}%
Charroin, Liza, Bernard Fortin, and Marie Claire Villeval. 2022. {``Peer
Effects, Self-Selection and Dishonesty.''} \emph{Journal of Economic
Behavior \& Organization} 200: 618--37.
\url{https://doi.org/10.2139/ssrn.3815848}.

\leavevmode\vadjust pre{\hypertarget{ref-chenCanSelfSelection2018a}{}}%
Chen, Roy, and Jie Gong. 2018. {``Can Self Selection Create
High-Performing Teams?''} \emph{Journal of Economic Behavior \&
Organization} 148 (April): 20--33.

\leavevmode\vadjust pre{\hypertarget{ref-coricelliPartnerSelectionPublic2004}{}}%
Coricelli, Giorgio, Dietmar Fehr, and Gerlinde Fellner. 2004. {``Partner
{Selection} in {Public Goods Experiments}.''} \emph{Journal of Conflict
Resolution} 48 (3): 356--78.

\leavevmode\vadjust pre{\hypertarget{ref-dahlDoesIntegrationChange2021}{}}%
Dahl, Gordon B, Andreas Kotsadam, and Dan-Olof Rooth. 2021. {``Does
{Integration Change Gender Attitudes}? {The Effect} of {Randomly
Assigning Women} to {Traditionally Male Teams}*.''} \emph{The Quarterly
Journal of Economics} 136 (2): 987--1030.

\leavevmode\vadjust pre{\hypertarget{ref-depaolaFreeridingKnowledgeSpillovers2019a}{}}%
De Paola, Maria, Francesca Gioia, and Vincenzo Scoppa. 2019.
{``Free-Riding and Knowledge Spillovers in Teams: {The} Role of Social
Ties.''} \emph{European Economic Review} 112 (February): 74--90.

\leavevmode\vadjust pre{\hypertarget{ref-esteves-sorensonGiftExchangeWorkplace2018}{}}%
Esteves-Sorenson, Constança. 2018. {``Gift {Exchange} in the
{Workplace}: {Addressing} the {Conflicting Evidence} with a {Careful
Test}.''} \emph{Management Science} 64 (9): 4365--88.

\leavevmode\vadjust pre{\hypertarget{ref-fehrTheoryFairnessCompetition1999}{}}%
Fehr, Ernst, and Klaus M. Schmidt. 1999. {``A {Theory} of {Fairness},
{Competition}, and {Cooperation}*.''} \emph{The Quarterly Journal of
Economics} 114 (3): 817--68.

\leavevmode\vadjust pre{\hypertarget{ref-felicianoStudentExperiencesUsing2016}{}}%
Feliciano, Joseph, Margaret-Anne Storey, and Alexey Zagalsky. 2016.
{``Student {Experiences Using GitHub} in {Software Engineering Courses}:
{A Case Study}.''} In \emph{2016 {IEEE}/{ACM} 38th {International
Conference} on {Software Engineering Companion} ({ICSE-C})}, 422--31.

\leavevmode\vadjust pre{\hypertarget{ref-fenollGenderMixTeam2022}{}}%
Fenoll, Ainoa Aparicio, and Sarah Zaccagni. 2022. {``Gender {Mix} and
{Team Performance}: {Differences} Between {Exogenously} and
{Endogenously Formed Teams}.''} \emph{Labour Economics} 79: 48.

\leavevmode\vadjust pre{\hypertarget{ref-fischerEffectsGermanUniversities2017}{}}%
Fischer, Mira, and Patrick Kampkötter. 2017. {``Effects of {German
Universities}' {Excellence Initiative} on {Ability Sorting} of
{Students} and {Perceptions} of {Educational Quality}.''} \emph{Journal
of Institutional and Theoretical Economics} 173 (4): 662.
\url{https://doi.org/10.1628/093245617X14816371560173}.

\leavevmode\vadjust pre{\hypertarget{ref-fischerWhenWhyTeams2023a}{}}%
Fischer, Mira, Rainer Michael Rilke, and B. Burcin Yurtoglu. 2023.
{``When, and Why, Do Teams Benefit from Self-Selection?''}
\emph{Experimental Economics}, March.

\leavevmode\vadjust pre{\hypertarget{ref-haaranenTeachingGitSide2015}{}}%
Haaranen, Lassi, and Teemu Lehtinen. 2015. {``Teaching {Git} on the
{Side}: {Version Control System} as a {Course Platform}.''} In
\emph{Proceedings of the 2015 {ACM Conference} on {Innovation} and
{Technology} in {Computer Science Education}}, 87--92. {ITiCSE} '15.
{New York, NY, USA}: {Association for Computing Machinery}.
\url{https://doi.org/10.1145/2729094.2742608}.

\leavevmode\vadjust pre{\hypertarget{ref-hamiltonTeamIncentivesWorker2003}{}}%
Hamilton, Barton H., Jack A. Nickerson, and Hideo Owan. 2003. {``Team
{Incentives} and {Worker Heterogeneity}: {An Empirical Analysis} of the
{Impact} of {Teams} on {Productivity} and {Participation}.''}
\emph{Journal of Political Economy} 111 (3): 465--97.
\url{https://doi.org/10.1086/374182}.

\leavevmode\vadjust pre{\hypertarget{ref-isomottonenChallengesConfusionsLearning2014}{}}%
Isomöttönen, Ville, and Michael Cochez. 2014. {``Challenges and
{Confusions} in {Learning Version Control} with {Git}.''} In
\emph{Information and {Communication Technologies} in {Education},
{Research}, and {Industrial Applications}}, edited by Vadim Ermolayev,
Heinrich C. Mayr, Mykola Nikitchenko, Aleksander Spivakovsky, and
Grygoriy Zholtkevych, 178--93. Communications in {Computer} and
{Information Science}. {Cham}: {Springer International Publishing}.
\url{https://doi.org/10.1007/978-3-319-13206-8_9}.

\leavevmode\vadjust pre{\hypertarget{ref-kiesslingSelfSelectionPeersPerformance2022}{}}%
Kiessling, Lukas, Jonas Radbruch, and Sebastian Schaube. 2022.
{``Self-{Selection} of {Peers} and {Performance}.''} \emph{Management
Science} 68 (11): 8184--8201.

\leavevmode\vadjust pre{\hypertarget{ref-knezFirmWideIncentives2001}{}}%
Knez, Marc, and Duncan Simester. 2001. {``Firm-{Wide Incentives} and
{Mutual Monitoring} at {Continental Airlines}.''} \emph{Journal of Labor
Economics} 19 (4): 743--72. \url{https://doi.org/10.1086/322820}.

\leavevmode\vadjust pre{\hypertarget{ref-luUsingPullBasedCollaborative2017}{}}%
Lu, Yao, Xinjun Mao, Gang Yin, Tao Wang, and Yu Bai. 2017. {``Using
{Pull-Based Collaborative Development Model} in {Software Engineering
Courses}: {A Case Study}.''} In \emph{Database {Systems} for {Advanced
Applications}}, edited by Zhifeng Bao, Goce Trajcevski, Lijun Chang, and
Wen Hua, 399--410. Lecture {Notes} in {Computer Science}. {Cham}:
{Springer International Publishing}.

\leavevmode\vadjust pre{\hypertarget{ref-masPeersWork2009}{}}%
Mas, Alexandre, and Enrico Moretti. 2009. {``Peers at {Work}.''}
\emph{American Economic Review} 99 (1): 112--45.

\leavevmode\vadjust pre{\hypertarget{ref-mcphersonBirdsFeatherHomophily2001}{}}%
McPherson, Miller, Lynn Smith-Lovin, and James M Cook. 2001. {``Birds of
a {Feather}: {Homophily} in {Social Networks}.''} \emph{Annual Review of
Sociology} 27 (1): 415--44.

\leavevmode\vadjust pre{\hypertarget{ref-ofek-shannyValidityMajorityMinorityPerformance2020}{}}%
Ofek-Shanny, Yuval. 2020. {``Validity of {Majority-Minority Performance
Gaps Measurements} on {PISA Tests}.''} SSRN Scholarly Paper ID 3670091.
{Rochester, NY}: {Social Science Research Network}.

\leavevmode\vadjust pre{\hypertarget{ref-pageMakingDifferenceApplying2007}{}}%
Page, Scott E. 2007. {``Making the {Difference}: {Applying} a {Logic} of
{Diversity}.''} \emph{Academy of Management Perspectives} 21 (4): 6--20.

\leavevmode\vadjust pre{\hypertarget{ref-ruefStructureFoundingTeams2003}{}}%
Ruef, Martin, Howard E. Aldrich, and Nancy M. Carter. 2003. {``The
{Structure} of {Founding Teams}: {Homophily}, {Strong Ties}, and
{Isolation} Among {U}.{S}. {Entrepreneurs}.''} \emph{American
Sociological Review} 68 (2): 195.

\leavevmode\vadjust pre{\hypertarget{ref-weidmannTeamPlayersHow2021}{}}%
Weidmann, Ben, and David J. Deming. 2021. {``Team {Players}: {How Social
Skills Improve Team Performance}.''} \emph{Econometrica} 89 (6):
2637--57.

\leavevmode\vadjust pre{\hypertarget{ref-wiseLowExamineeEffort2005}{}}%
Wise, Steven L., and Christine E. DeMars. 2005. {``Low {Examinee Effort}
in {Low-Stakes Assessment}: {Problems} and {Potential Solutions}.''}
\emph{Educational Assessment} 10 (1): 1--17.

\leavevmode\vadjust pre{\hypertarget{ref-yanComparingRegressionCoefficients2013}{}}%
Yan, Jun, Robert H. Aseltine, and Ofer Harel. 2013. {``Comparing
{Regression Coefficients Between Nested Linear Models} for {Clustered
Data With Generalized Estimating Equations}.''} \emph{Journal of
Educational and Behavioral Statistics} 38 (2): 172--89.

\end{CSLReferences}

\clearpage
\newpage

\section{Appendix}

\subsection{Appendix A}

This repository has 4 (5 for some) folders, named somewhat oddly. These
folders contain the project elaborations of the groups that have been
assigned to you.

We always anonymize the group names for us, and you will also get the
anonymized group names (naming of the folders) here. The anonymization
is done according to a particular system and is done anew for each
project.

Besides the project elaborations of the groups, a file with the ending
``\_Loesung.html'' is contained, which contains the sample solution for
this project. You should use this sample solution as a basis for your
evaluation of this project.

The focus of the evaluation is on the description of the tables and
graphs and, where required, their interpretation. However, it is also
important to scale graphics and tables sensibly and present them nicely
in this assignment. For example, there would be a deduction if the code
chunks in the group elaboration are visible in the HTML.

Please rate the written submissions and screencasts you receive on a
scale of 1-7 in the file TutorRating.Rmd, which is contained in the
respective folder per group, and knit the file afterward! Please
continue to write down in bullet points what led to point deductions in
your evaluation. These bullet points will then be played back to the
group along with the students' reviews. This will give the group some
more in-depth feedback and help them better categorize the reviews of
their fellow students.

\clearpage
\newpage

\subsection{Appendix B}

1.) In this course I learn things that fill me with enthusiasm (Likert
Scale)

2.) I did understand the most important topics of this project. (Likert
Scale)

3.) How satisfied are you with your team? (Likert Scale)

4.) How efficient was the teamwork in this project? (Likert Scale)

5.) How large do you think is your contribution to the project? (in \%)

6.) How evenly was the work distributed in your team? (Likert Scale)

7.) Do you think it is fair to get one grade per team? (Only after last
project)

8.) Which best describes your relationship to your group members? (Only
after first project):

\begin{itemize}
\item[$\square$]
  I did not know my group members before
\item[$\square$]
  I knew one or both of my group members before, but did not have much
  contact
\item[$\square$]
  One of the group members is a friend
\item[$\square$]
  Both group members are friends
\end{itemize}

\clearpage
\newpage

\subsection{Appendix C}

\begin{table}[H]

\caption{\label{tab:Balancetable}Balance Table}
\centering
\fontsize{8}{10}\selectfont
\begin{tabu} to \linewidth {>{\raggedright\arraybackslash}p{4cm}>{\raggedright}X>{\raggedright}X>{\raggedright}X>{\raggedright}X>{\raggedright}X>{\raggedright}X>{\raggedright}X}
\toprule
Variable & N & Mean & SD & N & Mean & SD & Test\\
\midrule
Treatment & Random &  &  & Self-selected &  &  & \\
gender & 124 &  &  & 121 &  &  & X2=0.167\\
... female & 41 & 33\% &  & 44 & 36\% &  & \\
... male & 83 & 67\% &  & 77 & 64\% &  & \\
hzbnote & 124 & 2.3 & 0.64 & 121 & 2.3 & 0.61 & F=0.068\\
\addlinespace
from\_local\_county & 124 &  &  & 121 &  &  & X2=0\\
... no & 99 & 80\% &  & 96 & 79\% &  & \\
... yes & 25 & 20\% &  & 25 & 21\% &  & \\
\bottomrule
\end{tabu}
\end{table}

\subsection{Appendix D}

\begin{table}[H] \centering    \caption{Effect of group-forming mechanism on GitHub-Contributions with Herfindahl-Index}    \label{}  \resizebox{0.98\textwidth}{!}{\begin{tabular}{@{\extracolsep{-5pt}}lccccc}  \\[-1.8ex]\hline  \hline \\[-1.8ex]   & \multicolumn{5}{c}{\textit{Dependent variable:}} \\  \cline{2-6}  \\[-1.8ex] & \multicolumn{5}{c}{\shortstack{Within-Group Herfindahl-Index of Share \\ of Code Words Committed}} \\  \\[-1.8ex] & (1) & (2) & (3) & (4) & (5)\\  \hline \\[-1.8ex]   Self-selection & 459.426 & 453.052 & 547.775 & 251.830 & 883.757 \\    & (377.884) & (381.043) & (382.418) & (497.452) & (537.714) \\    & & & & & \\   Best Group GPA &  & 100.412 & $-$149.749 & 87.565 &  \\    &  & (420.522) & (797.332) & (783.982) &  \\    & & & & & \\   Second Best Group GPA &  &  & 495.197 & 164.322 &  \\    &  &  & (620.764) & (609.718) &  \\    & & & & & \\   Max GPA Difference &  &  & 265.579 & 107.675 &  \\    &  &  & (534.121) & (558.906) &  \\    & & & & & \\   All Members Female &  &  &  & 650.797 &  \\    &  &  &  & (945.449) &  \\    & & & & & \\   One Member Male &  &  &  & 224.211 &  \\    &  &  &  & (543.858) &  \\    & & & & & \\   All Members Male &  &  &  & 863.846$^{**}$ &  \\    &  &  &  & (406.554) &  \\    & & & & & \\   Two Members Same Region &  &  &  & 320.304 &  \\    &  &  &  & (496.845) &  \\    & & & & & \\   Group Mean Test Exam &  &  &  &  & 2.276 \\    &  &  &  &  & (20.803) \\    & & & & & \\   Self-selection:Summer Term &  &  &  &  & $-$947.800 \\    &  &  &  &  & (790.583) \\    & & & & & \\  \hline \\[-1.8ex]  Project FE & Y & Y & Y & Y & N \\  Term FE & Y & Y & Y & Y & Y \\  Observations & 150 & 150 & 150 & 150 & 150 \\  R$^{2}$ & 0.033 & 0.034 & 0.052 & 0.095 & 0.049 \\  Adjusted R$^{2}$ & 0.006 & 0.0001 & 0.005 & 0.023 & 0.023 \\  Residual Std. Error & 1,690.396 & 1,695.578 & 1,691.241 & 1,676.316 & 1,676.043 \\  \hline  \hline \\[-1.8ex]  \multicolumn{6}{l}{$^{*}$p$<$0.1; $^{**}$p$<$0.05; $^{***}$p$<$0.01} \\  \multicolumn{6}{l}{\parbox[t]{14cm}{ Notes: Standard errors reported in parentheses are clustered at the group level. In the regressions, a group and a project define an observation. For the dependent variable we calculate the share of words from the number of words of code and text a student has contributed via GitHub to the total amount of words of code and text of her or his group in a project. We then calculate for each group and project the gini-coefficient of the shares of words of the group members, which is our dependent variable in these regressions. Best Group GPA is a control variable for the best high school GPA in a group, and Second Best Group GPA for the second best high school GPA in a group. German high school GPAs range from 1 to 4, where 1 refers to the best grade and 4 to the worst. Max GPA Difference is a control variable for the difference in high school GPA between the member of the group with the best GPA and the one with the worst GPA. Two members same region is 1 if two members of a group graduated from high school in the same or neighboring counties, otherwise it is 0.}} \\  \end{tabular}}  \end{table}

\clearpage
\newpage

\subsection{Appendix E}

\begin{table}[H] \centering 
  \caption{Effect of difficulty of task and relative skill level on contribution} 
  \label{} 
\begin{tabular}{@{\extracolsep{5pt}}lcc} 
\\[-1.8ex]\hline 
\hline \\[-1.8ex] 
 & \multicolumn{2}{c}{\textit{Dependent variable:}} \\ 
\cline{2-3} 
\\[-1.8ex] & \multicolumn{2}{c}{Share of GitHub Contributions} \\ 
 & \multicolumn{2}{c}{(Self-select)} \\ 
\hline \\[-1.8ex] 
 Own is Best Group GPA & 0.101$^{**}$ & 0.209$^{***}$ \\ 
  & (0.045) & (0.052) \\ 
  & & \\ 
 Hard Question & 0.013 & $-$0.012 \\ 
  & (0.033) & (0.029) \\ 
  & & \\ 
 Own is Second Best Group GPA & 0.089$^{*}$ & 0.081$^{**}$ \\ 
  & (0.050) & (0.037) \\ 
  & & \\ 
 Own is Best Group GPA:Hard Question & $-$0.026 & 0.071 \\ 
  & (0.047) & (0.045) \\ 
  & & \\ 
 Own is Second Best Group GPA:Hard Question & $-$0.047 & $-$0.013 \\ 
  & (0.044) & (0.038) \\ 
  & & \\ 
\hline \\[-1.8ex] 
Skill Controls & X & X \\ 
Home Region Controls & X & X \\ 
Gender Controls & X & X \\ 
Term FE & X & X \\ 
Project FE & X & X \\ 
Observations & 2,117 & 2,332 \\ 
R$^{2}$ & 0.052 & 0.084 \\ 
Adjusted R$^{2}$ & 0.045 & 0.078 \\ 
Residual Std. Error & 0.364 & 0.341 \\ 
\hline 
\hline \\[-1.8ex] 
\multicolumn{3}{l}{$^{*}$p$<$0.1; $^{**}$p$<$0.05; $^{***}$p$<$0.01} \\ 
\multicolumn{3}{l}{\parbox[t]{14cm}{ Notes: Standard errors reported in parentheses are clustered at the group level. In the regressions, a student and the task of a project define an observation. The dependent variable is the share of words contributed by a student to a project task. Own is Best Group GPA and Own is Second Best Group GPA indicate whether a student has the best or the second best high school GPA in their group. Best Group GPA is a control variable for the best high school GPA in a group, and Second Best Group GPA for the second best high school GPA in a group. German high school GPAs range from 1 to 4, where 1 refers to the best grade and 4 to the worst. Max GPA Difference is a control variable for the difference in high school GPA between the member of the group with the best GPA and the one with the worst GPA. Two members same region is 1 if two members of a group graduated from high school in the same or neighboring counties, otherwise it is 0. Individual With Best GPA in Group is 1 if the corresponding student has the best high school GPA in the group, otherwise it is 0.}} \\ 
\end{tabular} 
\end{table}

\clearpage
\newpage

\subsection{Appendix F}

\begin{figure}
\centering
\includegraphics{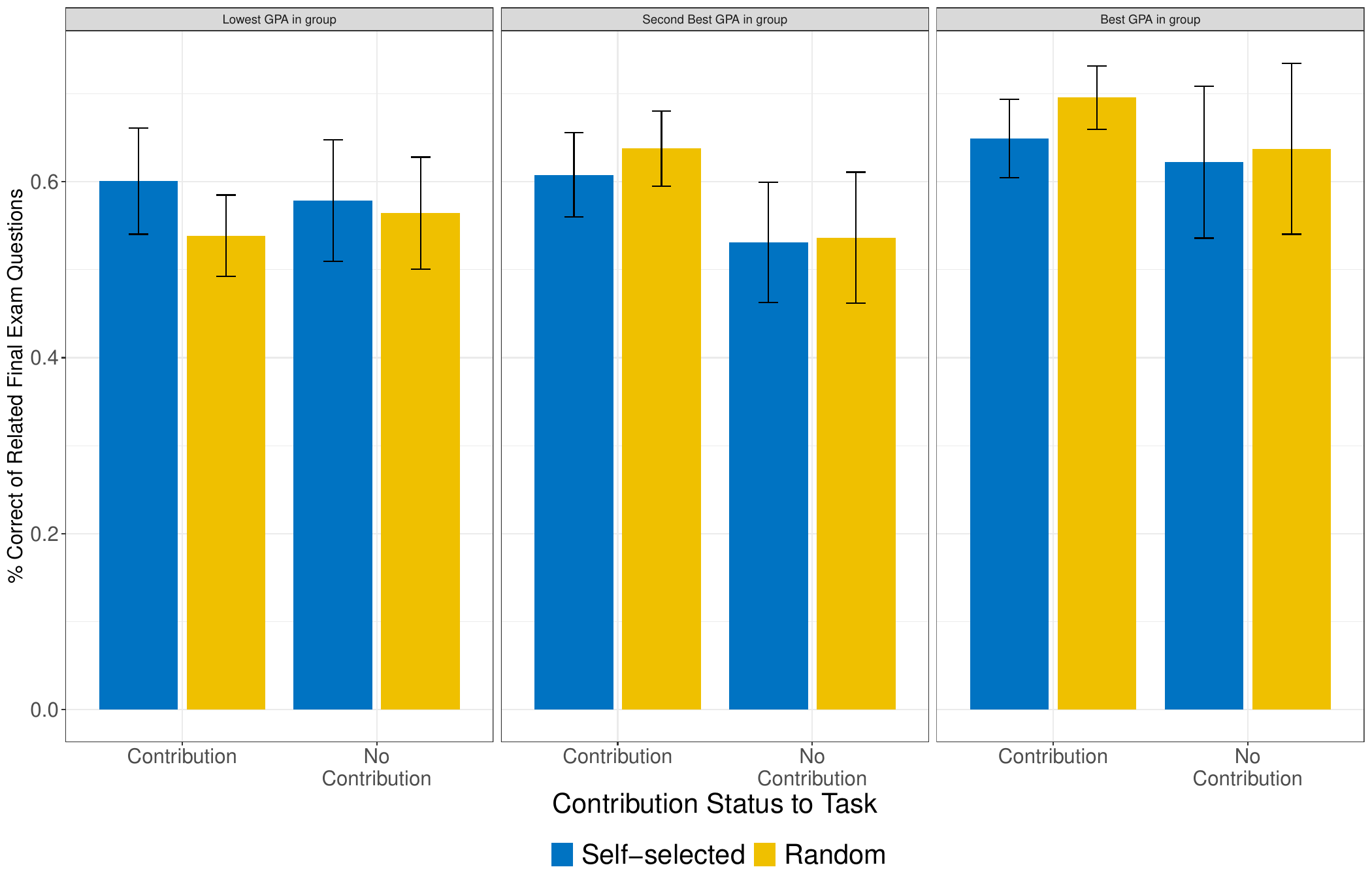}
\caption{Relationship between Skill, Effort Provision in Projects and
Knowledge Gain}
\end{figure}

\vspace{-1cm}

\begin{singlespacing}
Notes: In the graph we show how contributing to a project task via GitHub is correlated with the probability of answering an exam question correctly that directly relates to that task. Here an observation represents a student and a project task. The three panels show the averages and 95\%-confidence intervals separately for students who have the worst, second best, and best high school GPA of their group.

\end{singlespacing}

\subsection{Appendix G}

\begin{table}[H] \centering 
  \caption{Effect of group-forming mechanism on term productivity} 
  \label{} 
\begin{tabular}{@{\extracolsep{-5pt}}lccccc} 
\\[-1.8ex]\hline 
\hline \\[-1.8ex] 
 & \multicolumn{5}{c}{\textit{Dependent variable:}} \\ 
\cline{2-6} 
\\[-1.8ex] & \multicolumn{5}{c}{Project Percentage Points Term} \\ 
\\[-1.8ex] & (1) & (2) & (3) & (4) & (5)\\ 
\hline \\[-1.8ex] 
 Self-selection & $-$5.990$^{**}$ & $-$4.710$^{*}$ & $-$5.115$^{*}$ & $-$4.930$^{*}$ & $-$8.132$^{**}$ \\ 
  & (2.822) & (2.741) & (2.705) & (2.876) & (3.454) \\ 
  & & & & & \\ 
 Best Group GPA &  & $-$10.220$^{***}$ & $-$7.744 & $-$7.703 &  \\ 
  &  & (3.287) & (6.832) & (7.443) &  \\ 
  & & & & & \\ 
 Second Best Group GPA &  &  & $-$3.835 & $-$3.450 &  \\ 
  &  &  & (4.887) & (5.051) &  \\ 
  & & & & & \\ 
 Max GPA Difference &  &  & $-$0.648 & 0.380 &  \\ 
  &  &  & (4.466) & (4.896) &  \\ 
  & & & & & \\ 
 All Members Female &  &  &  & $-$0.082 &  \\ 
  &  &  &  & (5.045) &  \\ 
  & & & & & \\ 
 One Member Male &  &  &  & $-$0.342 &  \\ 
  &  &  &  & (3.751) &  \\ 
  & & & & & \\ 
 All Members Male &  &  &  & $-$3.039 &  \\ 
  &  &  &  & (4.025) &  \\ 
  & & & & & \\ 
 Two Members Same Region &  &  &  & 0.428 &  \\ 
  &  &  &  & (2.678) &  \\ 
  & & & & & \\ 
 Group Mean Test Exam &  &  &  &  & 0.360$^{**}$ \\ 
  &  &  &  &  & (0.171) \\ 
  & & & & & \\ 
 Self-selection:Summer Term &  &  &  &  & 3.833 \\ 
  &  &  &  &  & (5.761) \\ 
  & & & & & \\ 
\hline \\[-1.8ex] 
Project FE &  &  &  &  &  \\ 
Term FE & X & X & X & X & X \\ 
Observations & 86 & 86 & 86 & 86 & 86 \\ 
R$^{2}$ & 0.074 & 0.194 & 0.205 & 0.215 & 0.120 \\ 
Adjusted R$^{2}$ & 0.052 & 0.165 & 0.156 & 0.122 & 0.076 \\ 
Residual Std. Error & 13.207 & 12.393 & 12.462 & 12.706 & 13.033 \\ 
\hline 
\hline \\[-1.8ex] 
\multicolumn{6}{l}{$^{*}$p$<$0.1; $^{**}$p$<$0.05; $^{***}$p$<$0.01} \\ 
\multicolumn{6}{l}{\parbox[t]{13cm}{ Notes: Standard errors reported in parentheses are clustered at the group level. In the regressions, a group and a semester define an observation. The dependent variable is the mean of points awarded to the groups throughout a semester by the student assistants transformed to a 0 to 100 scale. Best Group GPA is a control variable for the best high school GPA in a group, and Second Best Group GPA for the second best high school GPA in a group. German high school GPAs range from 1 to 4, where 1 refers to the best grade and 4 to the worst. Max GPA Difference is a control variable for the difference in high school GPA between the member of the group with the best GPA and the one with the worst GPA. Two members same region is 1 if two members of a group graduated from high school in the same or neighboring counties, otherwise it is 0.}} \\ 
\end{tabular} 
\end{table}

\subsection{Appendix H}

\begin{table}[H] \centering    \caption{Effect of group-forming mechanism on productivity with lab controls}    \label{}  \resizebox{0.65\textwidth}{!}{\begin{tabular}{@{\extracolsep{-5pt}}lcccc}  \\[-1.8ex]\hline  \hline \\[-1.8ex]   & \multicolumn{4}{c}{\textit{Dependent variable:}} \\  \cline{2-5}  \\[-1.8ex] & \multicolumn{4}{c}{Project Percentage Points} \\  \\[-1.8ex] & (1) & (2) & (3) & (4)\\  \hline \\[-1.8ex]   Self-selection & $-$5.942$^{**}$ & $-$6.011$^{**}$ & $-$5.664$^{*}$ & $-$7.318$^{*}$ \\    & (2.881) & (2.806) & (2.875) & (3.801) \\    & & & & \\   Team Player in Group &  & 4.637$^{*}$ & 4.131 & 4.194 \\    &  & (2.737) & (2.890) & (2.738) \\    & & & & \\   Member Giving in Dictator Game &  & 6.707 & 1.006 & 5.801 \\    &  & (5.048) & (5.210) & (5.490) \\    & & & & \\   Altruist in Group &  & 6.177$^{*}$ & 6.446$^{*}$ & 6.108 \\    &  & (3.661) & (3.769) & (3.875) \\    & & & & \\   Conditional Cooperator in Group &  & 3.128 & 3.757 & 3.093 \\    &  & (3.182) & (3.238) & (3.181) \\    & & & & \\   Best Group GPA &  &  & $-$5.100 &  \\    &  &  & (7.446) &  \\    & & & & \\   Second Best Group GPA &  &  & $-$5.470 &  \\    &  &  & (5.080) &  \\    & & & & \\   Max GPA Difference &  &  & $-$0.289 &  \\    &  &  & (4.821) &  \\    & & & & \\   All Members Female &  &  & $-$0.631 &  \\    &  &  & (5.713) &  \\    & & & & \\   One Member Male &  &  & $-$0.559 &  \\    &  &  & (4.316) &  \\    & & & & \\   All Members Male &  &  & $-$3.555 &  \\    &  &  & (4.086) &  \\    & & & & \\   Two Members Same Region &  &  & $-$0.068 &  \\    &  &  & (2.799) &  \\    & & & & \\   Group Mean Test Exam &  &  &  & 0.259 \\    &  &  &  & (0.167) \\    & & & & \\   Self-selection:Summer Term &  &  &  & 2.127 \\    &  &  &  & (5.882) \\    & & & & \\  \hline \\[-1.8ex]  Project FE & X & X & X & X \\  Term FE & X & X & X & X \\  Observations & 166 & 166 & 166 & 166 \\  R$^{2}$ & 0.077 & 0.128 & 0.205 & 0.115 \\  Adjusted R$^{2}$ & 0.054 & 0.084 & 0.126 & 0.069 \\  Residual Std. Error & 15.958 & 15.707 & 15.345 & 15.830 \\  \hline  \hline \\[-1.8ex]  \multicolumn{5}{l}{$^{*}$p$<$0.1; $^{**}$p$<$0.05; $^{***}$p$<$0.01} \\  \multicolumn{5}{l}{\parbox[t]{12cm}{ Notes: Standard errors reported in parentheses are clustered at the group level. In the regressions, a group and a project define an observation. The dependent variable is the points awarded to the groups by the student assistants transformed to a 0 to 100 scale. Team Player in Group is a dummy variable indicating whether at least one group member scores 75 percent or more on the RMET. Member Giving in Dictator Game is a dummy variable, which is one when at least one group member is not keeping all the money in the dictator game. Altruist in Group and Conditional Cooperator in Group are two dummy variables computed from the prisoners dilemma, indicating whether there is at least one group member in a group, that cab be classified as altruist or conditional cooperator, respectively. Best Group GPA is a control variable for the best high school GPA in a group, and Second Best Group GPA for the second best high school GPA in a group. German high school GPAs range from 1 to 4, where 1 refers to the best grade and 4 to the worst. Max GPA Difference is a control variable for the difference in high school GPA between the member of the group with the best GPA and the one with the worst GPA. Two members same region is 1 if two members of a group graduated from high school in the same or neighboring counties, otherwise it is 0.}} \\  \end{tabular}}  \end{table}

\clearpage
\newpage

\end{document}